\documentclass[pre,floatfix,superscriptaddress,twocolumn]{revtex4-1}
\usepackage{graphicx}
\usepackage{subfigure}
\usepackage{bm}
\usepackage{amsmath}
\usepackage{lineno}
\usepackage{amssymb}
\usepackage{lipsum}
\usepackage{appendix}
\usepackage{color}
\usepackage{float}
\usepackage{soul}

\begin{document}

\title{Dynamical phase separation on rhythmogenic neuronal networks}

\author{Mihai Bibireata}
\affiliation{Department of Physics and Astronomy, UCLA, Los Angeles California, 90095-1596, USA}

\author{Valentin M. Slepukhin}
\affiliation{Department of Physics and Astronomy, UCLA, Los Angeles California, 90095-1596, USA}

\author{Alex J. Levine}
\affiliation{Department of Physics and Astronomy, UCLA, Los Angeles California, 90095-1596, USA}
\affiliation{Department of Chemistry and Biochemistry, UCLA, Los Angeles California, 90095-1596, USA}
\affiliation{Department of Computational Medicine, UCLA, Los Angeles California, 90095-1596, USA}

\begin{abstract}
We explore the dynamics of the preB\"{o}tzinger complex, the mammalian central pattern generator with $N \sim 10^3$ neurons, which produces a collective metronomic signal that times the inspiration. Our analysis is based on a simple firing-rate model of excitatory neurons with dendritic adaptation (the Feldman Del Negro model [Nat. Rev. Neurosci. 7, 232 (2006), Phys. Rev. E 2010 :051911]) interacting on a fixed, directed Erd\H{o}s-R\'{e}nyi network. In the all-to-all coupled variant of the model, there is spontaneous symmetry breaking in which some fraction of the neurons become stuck in a high firing-rate state, while others become quiescent. This separation into firing and non-firing clusters persists into more sparsely connected networks, and is partially determined by $k$-cores in the directed graphs. The model has a number of features of the dynamical phase diagram that violate the predictions of mean-field analysis. In particular, we observe in the simulated networks that stable oscillations do not persist in the large-N limit, in contradiction to the predictions of mean-field theory. Moreover, we observe that the oscillations in these sparse networks are remarkably robust in response to killing neurons, surviving until only $\approx 20 \%$ of the network remains. This robustness is consistent with experiment.    
\end{abstract}

\pacs{XXX}

\date{\today}

\maketitle


\section{Introduction}
\label{sec:Introduction}

Networks of connected neurons, or neuronal microcircuits, play a variety of roles~\cite{Koch:04}. Their collective dynamical properties depend upon both their 
network structure, {\em i.e.}, the pattern of which neurons are synaptically coupled to each other, and the neurons' individual properties, {\em i.e.}, 
the relationship between  input and output at the level of a single neuron. The preB\"{o}tzinger complex (preB\"{o}tC) is a particular neural microcircuit 
consisting of $ \sim 10^3$ neurons that acts as a central pattern generator, establishing the 
inspiratory rhythm in mammals~\cite{Smith:91,Bressloff:05,DelNegro2002}.  It is particularly interesting for theoretical study since it may be modeled as a network of 
essentially identical neurons that produces well-characterized
collective dynamics -- a metronomic oscillation between states of high firing rate and low firing rate, which sets the inspiratory rhythm. 

Feldman and Del Negro~\cite{DelNegro2002} suggested a simple model of the PreB\"{o}tC neurons that includes a combination of excitatory interactions and 
dendritic adaption, which was studied theoretically by Schwab {\em et al.}~\cite{Schwab:10} using a simple leaky integrate-and-fire model for the neurons~\cite{Koch:04,Morgado:08}. 
The principal idea enabling an oscillatory phase of the network  is 
the introduction of a slow internal variable (identified by experiment to be dendritic calcium concentration~\cite{Morgado:08})  in each neuron controlling 
its dendritic sensitivity to excitatory postsynaptic potentials (EPSP). The calcium concentration increases with each EPSP.  
In the proposed model, when the calcium concentration is below a threshold, the neurons 
are sensitive to external voltage signals (EPSPs) and, though mutual excitatory interactions, they collectively increase their firing rate. This collective
high-firing-rate period represents the inspiratory signal.  When 
many neurons have a high firing rate, however, the 
total input voltage to a typical neuron becomes sufficiently high that its dendritic calcium concentration rises above the threshold rendering that neuron insensitive 
to further input. As a result the neurons' firing rate rapidly decreases and remains small until the next period of mutual excitation that occurs once the 
dendritic calcium concentration of a typical neuron has fallen below the threshold restoring those neurons' sensitivity to input.  
 
In addition to the stably oscillating phase of the network, the model was shown to admit two other phases characterized by either steady-state low-firing rate or 
high firing-rate throughout the network. We refer to these as the quiescent and high-activity phases respectively.   A dynamical phase diagram of this system 
was obtained as a function of network size and basal (low calcium) neuronal excitability both in a mean-field analysis and numerically on a set of Erd\H{o}s -R\'{e}nyi (ER) graphs.  Intriguingly, the numerically obtained phase boundary between the stably oscillating and high activity phases demonstrates significant deviations 
from the mean-field theory predictions, with discontinuous ``jumps" whose position on the phase diagram corresponds to numbers of neurons at which the highest $k$
$k$-core~\cite{Dorogovtsev:06} of the network vanishes~\cite{Schwab:10}. 
 
In this manuscript, we further explore the Feldman-Del Negro model, showing that the dynamical phase space is much richer than previously thought. 
In particular, we find that region of phase space consistent with stable oscillations is bounded in both network size and neuronal basal excitability.  This 
is inconsistent with the mean-field predictions.  Moreover, most of the phase boundaries are jagged (the interfaces are rough), which is also 
 inconsistent with the mean-field predictions for this dynamical system.  
The main reason for the failure of the mean-field approximation is due to the spontaneous separation of the neurons into groups with disparate firing rates. In fact, 
even in an all-to-all coupled network, the permutation symmetry of the neurons is spontaneously broken into these high and low firing-rate groups. We call this spontaneous activity phase separation.
We analyze this separation process on random networks both numerically and analytically, showing that the connectivity disorder of the random networks guides 
the separation process. 

We organize the rest of the manuscript as follows. In section~\ref{sec:Nonmon}, we demonstrate spontaneous 
activity separation into high and low firing-rate groups on a small network. We  
 show how that separation leads to the observed roughness of the phase boundaries. For the special case of an all-to-all coupled network, one can analytically derive activity phase separation. We do so and compare these results to numerical simulations on all-to-all coupled networks in section~\ref{sec:fully}.  From that analysis we learn that 
the steepness of the neuronal firing rate function (as a function of somatic potential) controls this spontaneous symmetry breaking on the network.  In section~\ref{sec:ER}, we move to the case of 
more sparsely connected networks, chosen from the ensemble of ER networks, where we prove that the activity separated solution, if it exists, is stable for sufficiently 
sharp neuronal firing-rate functions.  The cases where such activity separated solutions do not exist is reminiscent of converse symmetry breaking~\cite{Nishikawa:16}, where symmetric 
solutions can be paradoxically stabilized by system asymmetry. Finally, in section~\ref{sec:k-cores} we consider the role of $k$-cores in determining which 
neurons fall into the high-activity state in sparsely connected networks.  We prove that, when setting the low somatic voltage firing rate to zero, activity phase separation is 
exactly controlled by the $k$-cores. We suggest that $k$-cores remain relevant in controlling the phase boundary between the quiescent and high activity phases of the disordered system, but these 
topological features cannot alone account the roughness of the high-activity/stable oscillation phase boundary. 
We conclude with a comparison of these results to current experiments and make suggestions for future work.  

The reader interested primarily in our experimental predictions for {\em in vitro}  PreB\"{o}tC system may turn to section~\ref{sec:Discussion}. Readers interested in the simulations will find the reference to the software and appropriate parameters in Appendix \ref{app:fig-parameters}. 

\section{The Feldman Del Negro model}
\label{sec:Nonmon}
Following Ref.~\cite{Schwab:10}, we describe the two-compartment neuron model of preB\"{o}tC neurons.  The $i^{\rm th}$ neuron  is characterized by two dynamical 
variables, its somatic potential $V_i$ and its dendritic calcium concentration $C_i$. Their dynamics are controlled by the equations
\begin{eqnarray}
    \label{eq:V}
    \frac{dV_i}{dt} &=& \frac{1}{\tau _V}(V_{eq}-V_i) + \Delta V(C_i) \sum _{j} M_{ij} r(V_j)\\
     \frac{dC_i}{dt} &=& \frac{1}{\tau _C}(C_{eq}-C_i) + \Delta C \sum _{j} M_{ij} r(V_j),
    \label{eq:C}
\end{eqnarray}
where $\Delta V(C)$ and $r(V)$ are defined by
\begin{equation}
\label{eq:VC}
\Delta V(C) =  \Delta V_{max}\sigma \left( \frac{C^*-C}{g_C} \right)
\end{equation}
and
\begin{equation}
\label{eq:rV}
 r(V) = (r_m - r_b) \sigma \left( \frac{V-V^*}{g_V} \right)  + r_b.
\end{equation}
In Eqs.~\ref{eq:VC}, \ref{eq:rV} we have introduced the standard sigmoid (Fermi) function
\begin{equation}
\sigma(x) = \frac{1}{1 + e^{-x} }. 
\label{eq:sigmoid}
\end{equation}
Here and throughout the manuscript we work in dimensionless calcium concentration units obtained by setting $C_{\rm eq}=0$ and $C^{*}=5$.

Eq. \ref{eq:V} is typical of a leaky integrate and fire model for an excitatory neuron. The principal addition in the two-compartment model is dendritic adaptation which is built into $\Delta V(C_i) $ defined in Eqs. \ref{eq:VC} and \ref{eq:sigmoid}. An incoming EPSP produces both an  increase in dendritic calcium concentration $\Delta C$ and somatic potential $\Delta V(C) $. But above a threshold concentration $C^*$, $\Delta V (C) $ becomes small, rendering the neuron insensitive to EPSPs. In the absence of incoming EPSPs, the dendritic calcium concentration returns to $C_{eq} $ on a time scale of $\tau_C$ at which point the neuron is once again sensitive to EPSPs. 

The parameter space of the neuron model is controlled by a small set of physiological constants. There are the steady-state dendritic calcium concentration and somatic potential $C_{eq}$ and $V_{eq}$ respectively. The voltage-dependent firing rate is determined by the basal and maximal firing rates $r_b$ and $r_m$ as well as $g_V$, which controls the steepness of the transitions around the threshold voltage $V^*$.  Dendritic adaptation is parametrized by the maximum voltage increment associated with an EPSP $\Delta V_{max} $, a calcium concentration threshold $C^*$ and a steepness parameter $g_C$, analogous to $g_V$ discussed above. In addition to the two time scales $\tau_V < \tau_C$ for the relaxation of somatic potential and dendritic calcium, there is a fixed calcium concentration increment $\Delta C$ associated with the response to an EPSP.  Table ~\ref{tab:parameters} provides the currently available values of the model parameters.


\begin{table}
\centering
\caption{Model parameters known from experiment}
\begin{tabular}{l|c|r}
\textbf{Parameter} & \textbf{Approximate value} & \textbf{References} \\
\hline \hline
$V_{eq}  $ &  -65 mV  & \cite{Koch:04}  
\\
$V^* $ & -50 mV & \cite{Koch:04}
\\
$\tau_V $ & 20 ms & \cite{Rekling2000},\cite{Ashhad:19}
\\
$r_m$ & 40 Hz & \cite{Rekling2000},\cite{Ashhad:19}
\\
$r_b$ & 0.1 Hz & \cite{Ashhad:19}
\\
$\Delta V_{max}  $ & 2.8 mV & \cite{Rekling2000},\cite{Ashhad:19} 
\\
$p$ & 0.065 & \cite{Rekling2000}
\\
$N$ & $10^3$ & \cite{Rekling2000}, \cite{Gray:01} 
\end{tabular}
\label{tab:parameters}
\end{table}

The model also depends on the size and connectivity of the underlying network of synaptic connections between the neurons.  The network's structure can be encoded by an adjacency 
matrix ${\cal M}$ whose matrix elements $M_{ij} = 1$ if neuron $i$ synapses on neuron $j$, and equal to zero otherwise. In this manuscript, we consider only networks built from uncorrelated 
stochastic connections -- Erd\H{o}s R\'enyi directed graphs~\cite{Erdos:59}. An ensemble of such networks is determined by a single probability $p$ that any non-diagonal matrix element is equal to one.  We exclude
the possibility of a neuron synapsing on itself.  The all-to-all network is simply  the case of such an ER graph with $p=1$.

\subsection{Dynamical phase diagram}

For a given set of parameters and given network of $N$ neurons, the dynamical system evolves deterministically from a set of $2N$ initial conditions leading to either a fixed point, limit cycle, or chaotic 
dynamics at long times.  We find multiple fixed points, which can be  further distinguished as quiescent (Q) where the somatic potential averaged over the network of 
neurons $\langle V \rangle$ lies below the transition to the high-firing state $V^*$, or high activity (HA), where $\langle V \rangle  > V^*$. Similarly, we can distinguish three classes of stable limit cycle oscillations: 
below threshold oscillations (BTO) where the oscillatory average voltage remains below the threshold for high firing rate, above threshold oscillations (ATO), where the oscillatory average 
voltage remains above the threshold for high firing rate, and true metronomic activity (TMA), where the stable limit cycle oscillations carry the system between high and low firing rates, producing the
physiologically observed inspiratory rhythm. 

\begin{figure}
    \centering
     \includegraphics[width=85mm]{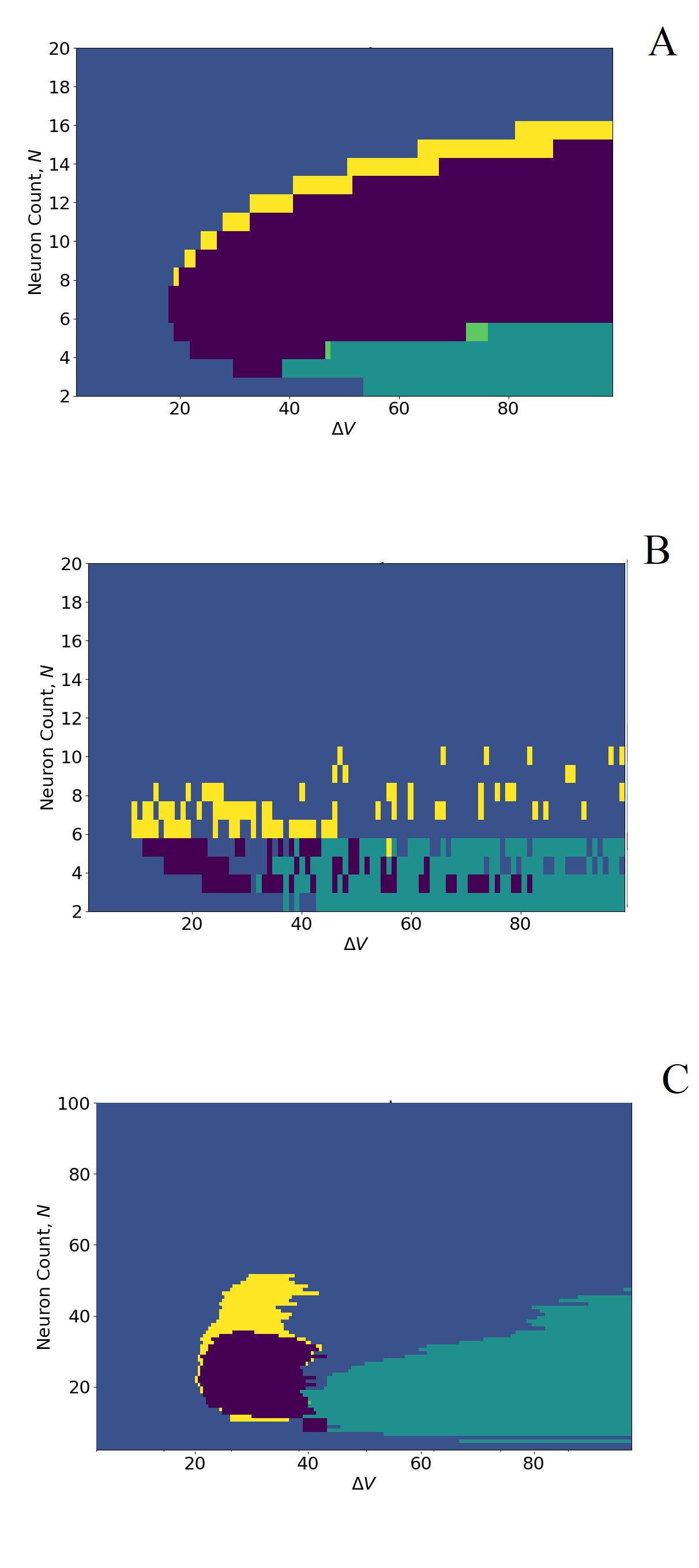}
     \caption{Dynamical phase diagram of the model as a function of the size of the network $N$ and basal neuronal excitability $\Delta V$.  (A) An all-to-all coupled network with large $g_{C} = 3$ produces
     phase behavior consistent with mean-field predictions, but (B) sharp sigmoids (small $g_{C} = 0.01$) produce a disordered diagram in which all dynamical phases are strongly mixed and the network
     dynamics is highly dependent on initial conditions. Finally, in (C) randomly connected networks ($p = 0.2$) with large $g_{C} = 3$, have initial-condition independent results with a modified 
     dynamical phase diagram. In all three panels the phases are: Q (blue), BTO (yellow), HA (dark green), ATO (light green), TMA (purple). All parameter values are listed in appendix~\ref{app:fig-parameters}.}
    \label{fig:general}
\end{figure}

To examine the dynamical phase behavior of the system, we vary the basal excitability of the neurons and the size of their network --  $\Delta V_{max} $ and $N$ -- while 
fixing the rest of the parameters. Typical results for networks 
are shown in Fig.~\ref{fig:general} for three different choices of the fixed variables.
We observe in Fig.~\ref{fig:general}A the numerically determined phase diagram for all-to-all coupled networks, which agrees with the mean field solution of the model shown in Appendix C. In general, we find that the numerically determined phase diagram agrees with the mean field approximation (for arbitrary initial conditions) as long as the transition in dendritic sensitivity is sufficiently smooth, {\em i.e.} 
$g_C \gtrsim  1$ (see Appendix \ref{sec:mean-field} ).  In this limit we observe all five dynamical phases: Q (blue), BTO (yellow), ATO (light green), TMA (purple) and HA (dark green).

There are two ways to invalidate the mean-field predictions.  The first is to make the dendritic calcium adaptation more abrupt, {\em i.e.}, decrease  $ g_C \lesssim 1 $ while retaining the all-to-all coupling. In that case, we encounter a much more complex phase space as shown in Fig.~\ref{fig:general}B where the phases mix on a small scale in the parameter space. We discuss analytically the quasi-periodic pattern arising in this case in the Appendix~\ref{app:fractal}. We also observe a dependence upon initial conditions. In effect, the dynamical phase diagram is not only highly heterogeneous, but also the regions that we associate with a particular phase may depend on the choice of initial conditions. The mean-field analysis is non-predictive, and one may say that the even the introduction of a dynamical phase diagram itself is not as well defined as in the mean-field case. 

The second way to invalidate the mean-field predictions is more interesting.  We maintain the smooth neuronal sigmoids, but reduce the number of network connections.  
In that case, as shown in  Fig.~\ref{fig:general}C, the phase behavior of the network is once again 
insensitive to initial conditions.  Moreover, the general
structure of the mean-field phase diagram is preserved, but the phase boundaries are distorted.  Both the HA (green) and Q (blue) phases expand, while the physiologically relevant TMA (purple) 
phase shrinks.   Both the TMA and BTO (yellow) phase are now bounded, whereas they extended to arbitrarily large $\Delta V$ in the mean-field prediction.  In this regime,
we do not see chaotic dynamics unlike in the cases where $g_{C}$ is small. Changing other parameters of the model changes the shape of the phase boundaries, but does not introduce 
new dynamical phases. Both routes to the breakdown of mean field theory (small $g_C$ and more sparsely connected networks) are related to an inherent instability toward activity phase separation. We discuss this in more detail below. 

\begin{figure}
    \centering
    \includegraphics[width=85mm]{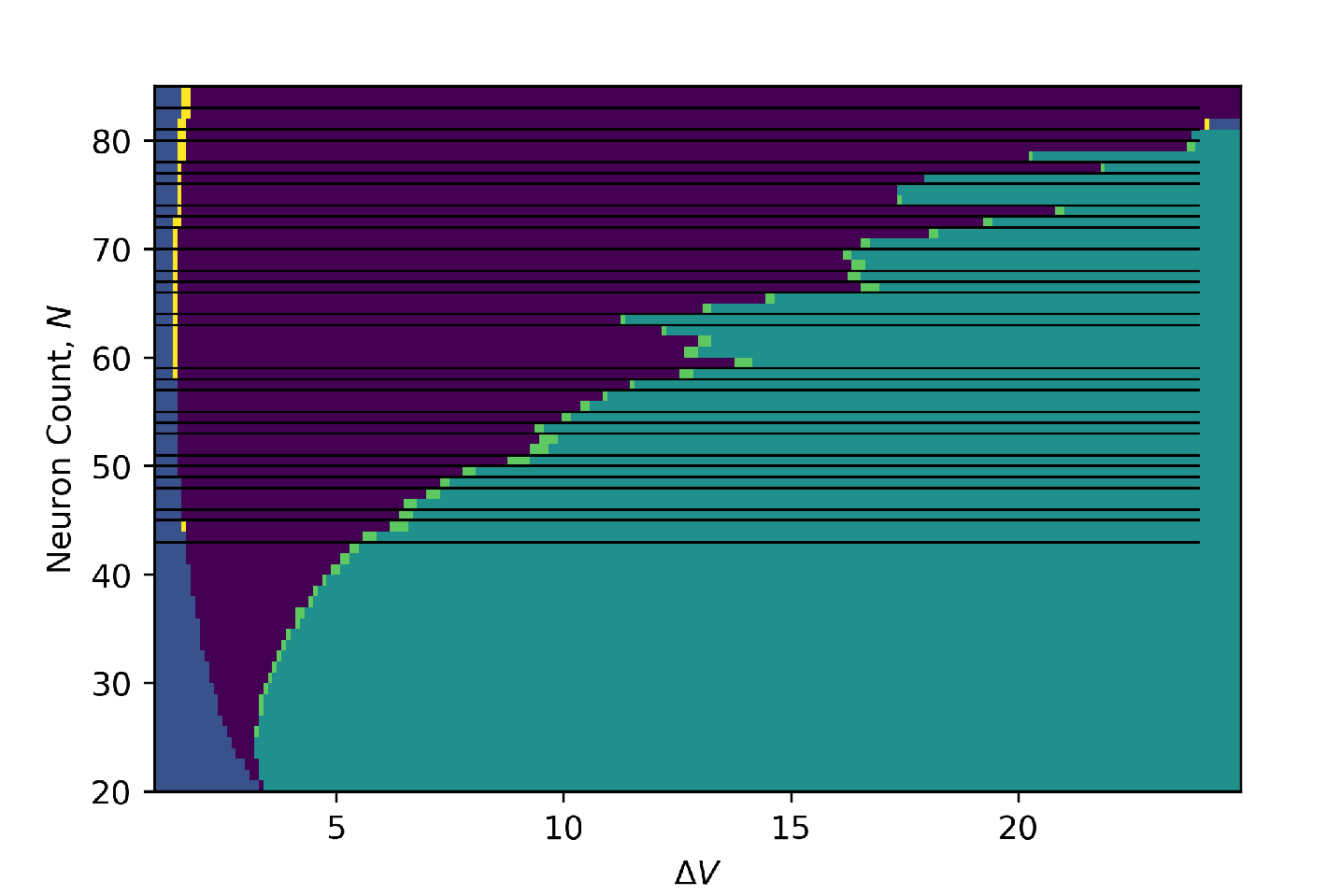}
    \caption{Reentrant behavior along the of the TMA-HA phase boundary. $k$-cores transition are shown as black lines, colors are the same as at the previous figure. }
    \label{fig:nonmon}
\end{figure}

Before discussing the phase separation, we note that the roughness of the phase diagram in the non-mean-field regime, as shown in Fig.~\ref{fig:general}C, implies that the 
physiologically desirable stably oscillating phase (TMA) admits a type of reentrant behavior in which one can remove neurons (decrease $N$) from a network in the high activity state to render it
in the stably oscillating TMA phase.   In Fig.~\ref{fig:nonmon} we see examples of such possible transitions at $\Delta V =18$mV  where by decreasing the number of neurons from $N = 90$ to 
$N=65$, one encounters two transitions HA to TMA, before remaining in the HA phase below $N =72$.  This suggests a specific experimental test of the fundamental model that can be made by 
looking for these reentrant dynamical transitions upon killing neurons in the network. 
In this figure we also show with black horizontal lines the values of $N$ at which various $k$-cores of the network vanish.  The positions of the tongues of extra stability of the oscillating TMA phase
appear to be bounded by these $k$-core transitions, suggesting that the disappearance high-$k$ $k$-cores changes the stability of the oscillatory (TMA)  phase.  We return to this point in section~\ref{sec:k-cores},  
where we show that $k$-cores play a dominant role in the phase stability of a somewhat simplified version of the model.

\begin{figure}
    \centering
    \includegraphics[width=85mm]{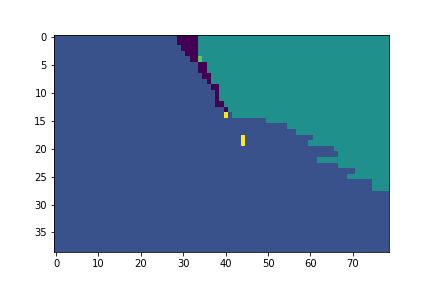}\\
    \caption{Phase diagram showing reentrant behavior at the Q (blue) HA (dark green) phase boundary.  There are also small regions of the oscillatory phases: TMA (purple), ATO (light green), BTO (yellow).  }
    \label{fig:nonmonotonicity}
\end{figure}

The fact that removing neurons from the network can enhance its ability to maintain stable oscillations seems to be counter-intuitive.  
This reentrant behavior appears at many phase boundaries in the system, including the one between the high-activity (HA) and quiescent (Q) phases.  An example of such reentrant behavior
at this phase boundary is shown in Fig.~\ref{fig:nonmonotonicity}. To understand how this 
behavior emerges, it is simpler to study this case where the neurons' dynamics reaches a fixed point rather than a limit cycle.  
Consider a fixed point of the system; setting the time derivatives on the left hand side of Eqs.~\ref{eq:V},~\ref{eq:C}, we 
obtain
\begin{eqnarray}
     V_i &=&  V_{eq} + \Delta V(C_i) \tau _V \sum _{j} M_{ij} r(V_j)\\
     C_i &=& C_{eq} + \Delta C \tau _C \sum _{j} M_{ij} r(V_j)
\end{eqnarray}
For neuron $i$ to be rapidly firing, it must receive a number of EPSPs consistent with both  $V_i > V^*$ and $C_i < C^* $.  In this way, its somatic voltage is maintained above the threshold and it remains
sensitive to EPSPs.  Too many EPSPs will drive $C_i > C^* $, resulting in the neuron's somatic potential falling below that threshold, while too few EPSPs will allow $V_i < V^*$ even while maintaining
dendritic sensitivity. As a result, the stable configuration of $V_i > V^*$ and $C_i < C^*$
can be destroyed by either adding or removing neurons that synapse on neuron $i$. We can see precisely how this works in an example of a small network of 
seventeen neurons poised near the HA-Q boundary.  

\begin{figure*}
    \centering
    \includegraphics[width=200mm]{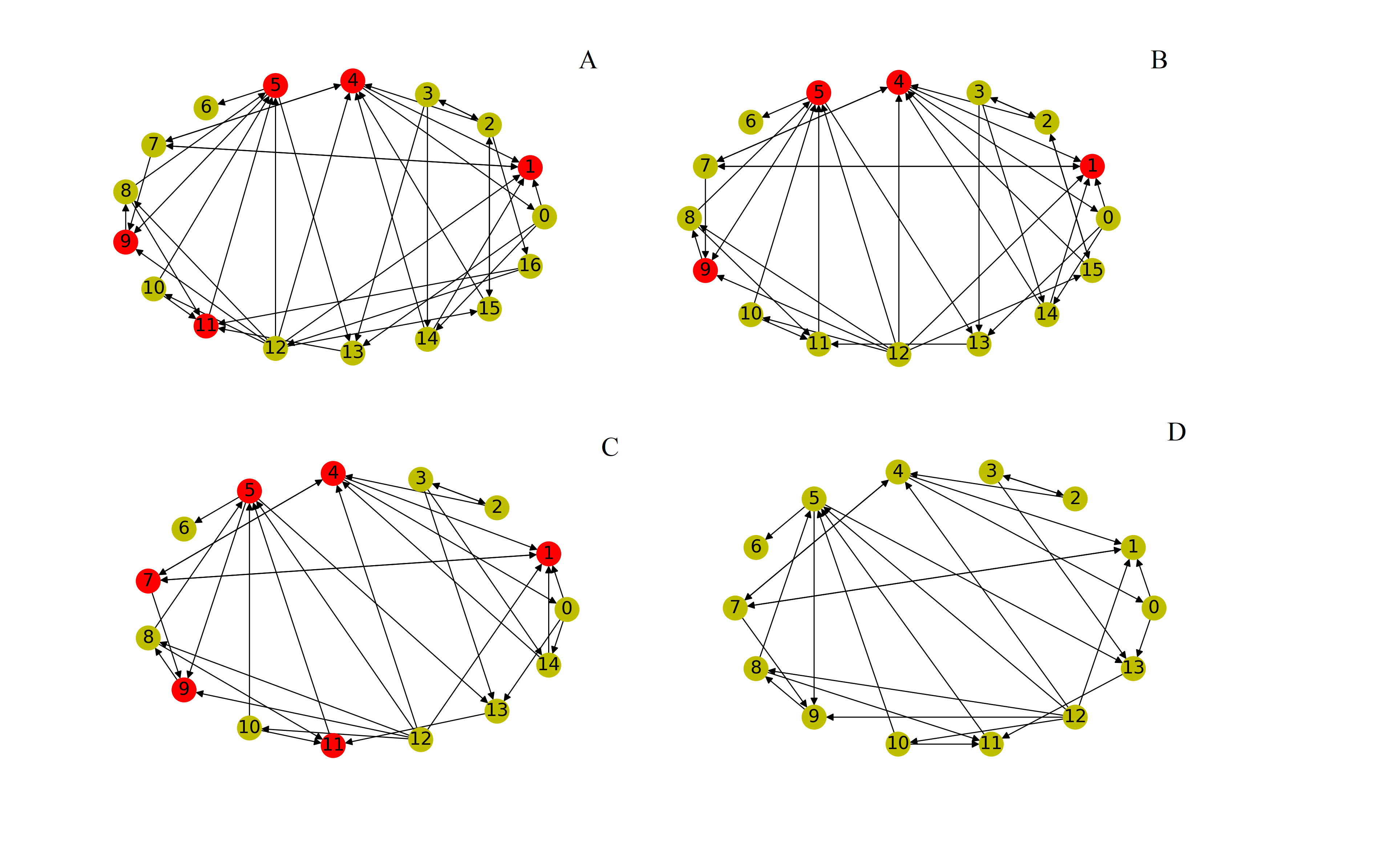}
    \caption{ Example of reentrant high activity on a small network. Red neurons have $V> V^{*}$ and yellow $V < V^*$.   The neurons are numbered and the last neuron in each network is removed when going $A \rightarrow B \rightarrow C \rightarrow D $. With the removal of neuron 16 (from A to B) , the somatic potential of the neuron 11 drops below the threshold, as it has insufficient voltage input, and the average network voltage falls below $V^*$, too. Going from B to C neuron 15, which synapses to neuron 4, is removed, what lowers its calcium concentration. As a consequence, the somatic potential of the neuron 4 increases as well as its firing rate, resulting in increasing of the firing rate and voltage input to the neuron 7. The somatic potential of the neuron 7 then goes above the threshold, too. The increasing firing rate of neuron 4 also raises the somatic potential of neuron 0, which raises somatic potential of neuron 13, which in turn raises it for the neuron 11. Although somatic potentials of neurons 0 and 13 does not exceed $V^*$, for neuron 11 it does. As a net result, the average voltage of the network raises above $V^*$.   Finally, when neuron 14 is removed (from C to D), all neurons are deactivated and $\Delta V$ must increase to restore high activity.
}
    \label{fig:four-networks}
\end{figure*}
In Fig.~\ref{fig:four-networks} we see that eliminating a low-firing rate neuron (number 16) from the network causes neuron 11 to change from high to low firing rate.  Removing an excitatory 
neuron has the  expected behavior of reducing the total activity of the network. But the subsequent removal of another low-firing rate neuron (neuron 15) results in neurons 7 and 11 once 
again returning to high firing rate.  Finally, by removing the low-firing rate neuron 14, the entire network collapses into the quiescent state.  

\section{Spontaneous symmetry breaking on all-to-all networks}
\label{sec:fully}

In this section we explore phase separation on all-to-all coupled networks, i.e., those having an adjacency matrix of the form $M_{i j } = 1$ for all $i \neq j$ and
$M_{ii}=0$ for all $i$.  The steady-state of the system spontaneously breaks the permutation symmetry of the neurons.  To explore this symmetry breaking, we 
first investigate the symmetry preserving solution, that is obtained from the pair of differential equations
\begin{eqnarray}
    \frac{dV}{dt} &=& \frac{1}{\tau_V}(V_{eq}-V) + (N - 1) \Delta V(C)  r(V)\\
     \frac{dC}{dt} &=& \frac{1}{\tau_C}(C_{eq}-C) + (N - 1)\Delta C   r(V),
\end{eqnarray}
which results from setting $C_{i}=C(t)$ and $V_{i}= V(t)$ for all $i$ in Eqs.~\ref{eq:V},~\ref{eq:C} and using the all-to-all adjacency matrix.  

We demonstrated numerically that the dynamics of the full system Eqs.~\ref{eq:V},~\ref{eq:C}  evolves towards this permutation symmetric solution for arbitrary 
initial conditions if the sigmoidal functions $P(V)$ and $\Delta V (C)$ are smooth enough.  If, on the other hand, these sigmoids are sharper, the system is unstable 
towards activity phase separation (breaking the original permutation symmetry of the underlying network) into time-independent 
subnetworks of high and low firing rate neurons when the initial conditions are not themselves identical across the 
network.  This symmetry broken state is, of course, not captured by the above mean field analysis.  To explore it, we need to analyze the full system of equations. 

\begin{figure}
    \centering
    \includegraphics[width=85mm]{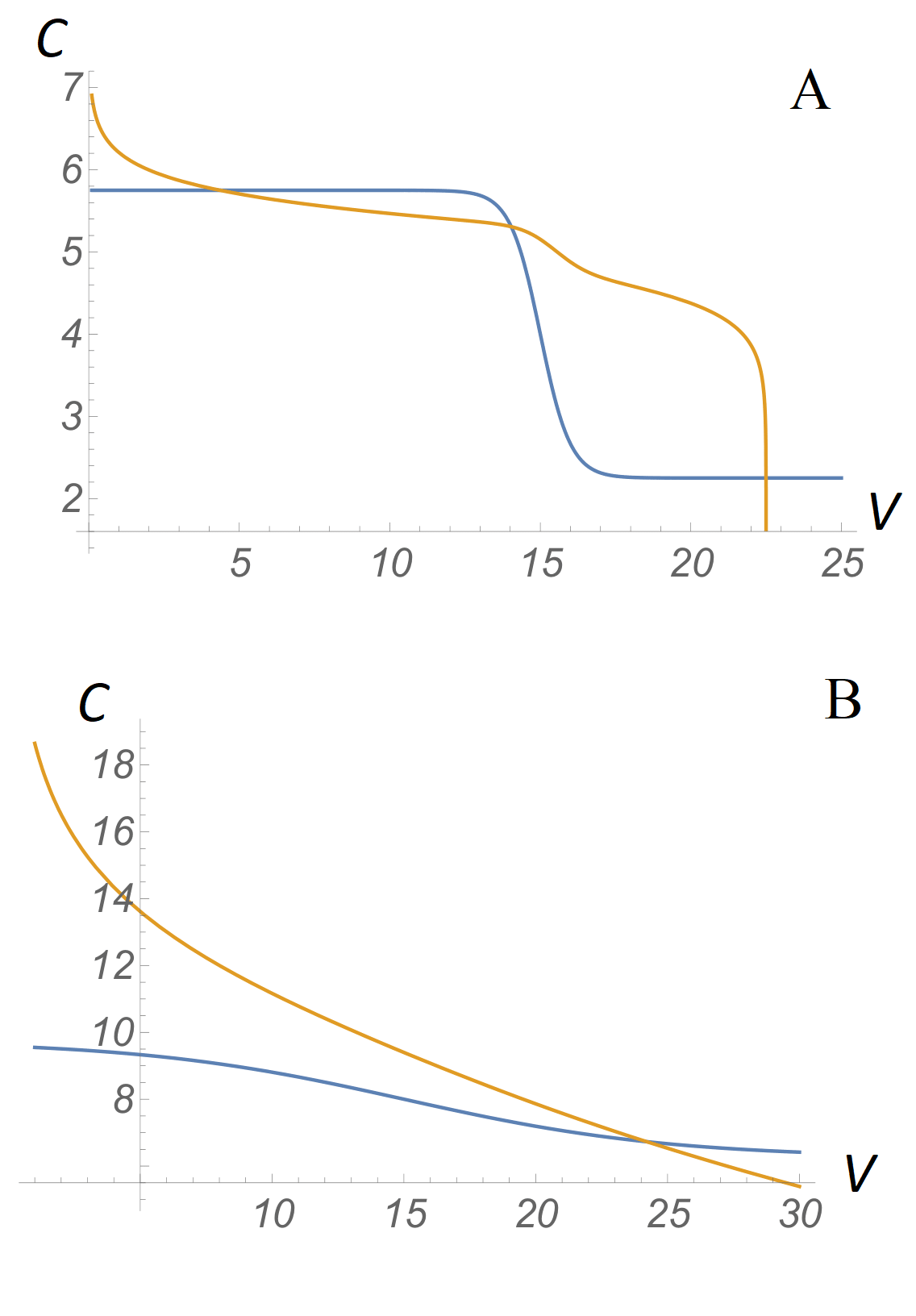}
    \caption{Nullclines of the all-to-all $N  = 10$ network described by Eqs.~\ref{eq:V-R} (orange) and \ref{eq:C-R} (blue) in the text. There are either three fixed points or one
    fixed point depending on parameters.  Assuming $R$ is constant (and not fixed self-consistently) two of the  fixed points annihilate in a standard pitchfork bifurcation~\cite{Strogatz:2000}. (A) $g_V = 0.5$ mV, $g_C = 0.3$, three fixed points. (B) $g_V = 5$ mV, $g_C = 3$, one fixed point. }
    \label{fig:nullclines}
\end{figure}

Defining the sum of firing rates over the entire network as $R = \sum_i r(V_i) $, we rewrite the dynamical system as 
\begin{eqnarray}
\label{eq:V-R}
    \frac{dV_i}{dt} &=& \frac{1}{\tau _V}(V_{eq}-V_i) + \Delta V(C_i) \left[ R - r(V_i) \right]\\
     \label{eq:C-R}
    \frac{dC_i}{dt} &=& \frac{1}{\tau _C}(C_{eq}-C_i) + \Delta C \left[ R - r(V_i) \right].
\end{eqnarray}
Now we can look for a self-consistent solution of this system, {\em i.e.}, we find $V_i(R) $ such that $R = \sum_i r(V_i) $. Studying the nullclines of
Eqs.~\ref{eq:V-R},~\ref{eq:C-R}, we see that it can have one to three fixed points. The cases of one and three fixed points are 
shown in Fig.~\ref{fig:nullclines}, where we see the intersections of the nullclines of Eqs.~\ref{eq:V-R} and \ref{eq:C-R} in orange and blue respectively. 

Since we are looking for a self-consistent solution, we can not analyze its stability directly from the graph; however, we can find the fixed points and later 
analyze their stability. 
For neuronal parameters consistent with smooth sigmoids, there is only one fixed point $V_f, C_f$ for a fixed value of $R$.  This self-consistent solution is both permutation symmetric and consistent with our mean field prediction.   
In contrast,  for sharp sigmoids, there is more than one fixed point, so it is possible to find some fraction of the network neurons at a high-voltage
fixed point, while the remainder is at a low-voltage fixed point. The number of neurons in these two categories is determined by the 
condition $R = \sum_i r(V_i) $. There is also a range of parameters with $g_V > 0, g_C = 0$, such that the self-consistent solution does not exist.Therefore, there is no fixed point, and only oscillations are allowed. See Appendix~\ref{app:fractal} for the details of the analytical calculation. 

\begin{figure*}
    \includegraphics[width=185mm]{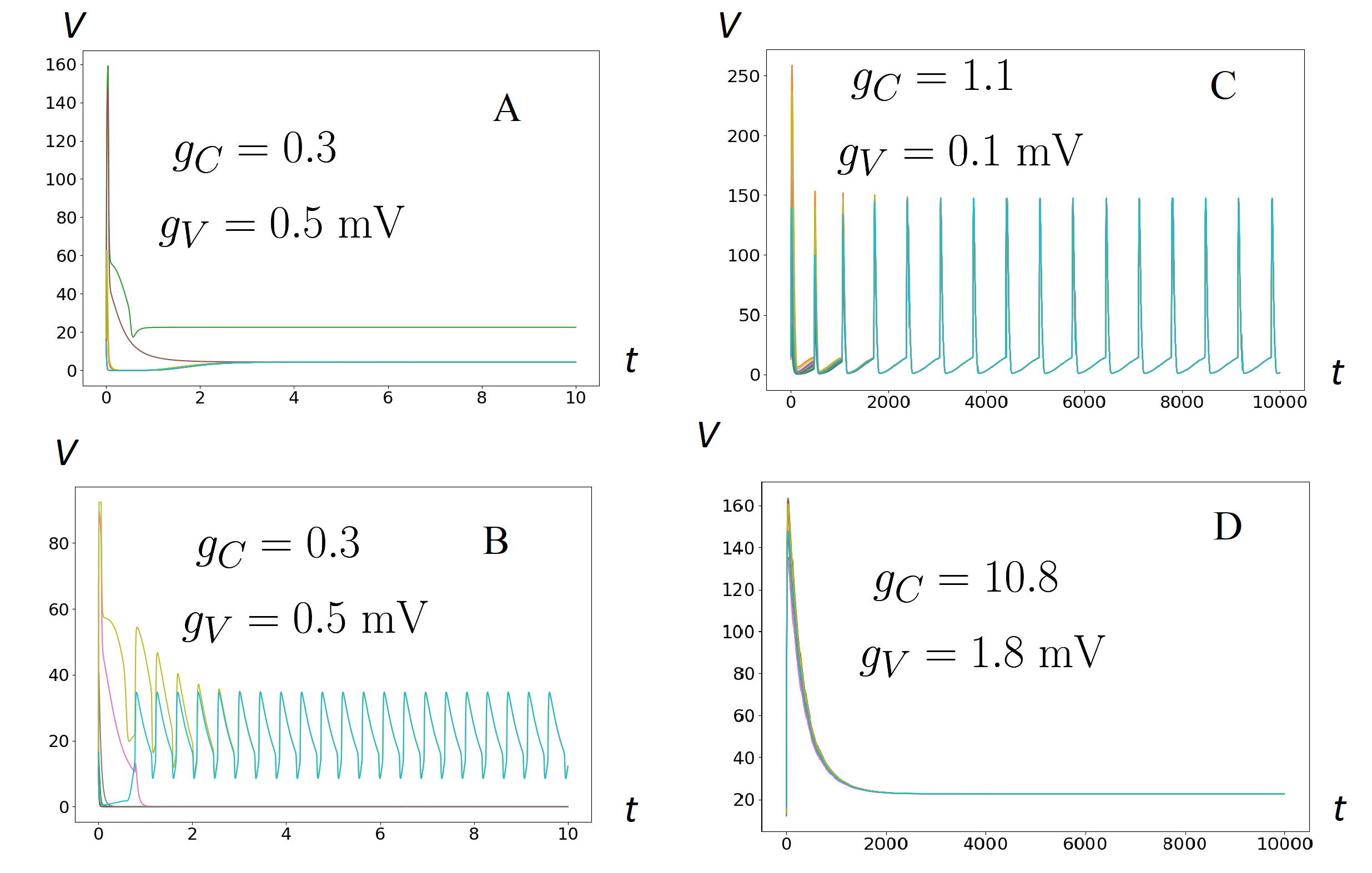}
    \caption{Activity phase separation on all-to-all connected network of $N = 10$ neurons. The traces show somatic potential of individual neurons as a function of time. Corresponding values of $g_C$ and $g_V$ are shown on the panels.   (A) One neuron is at high voltage , nine are quiescent.  (B) Two neurons oscillate,  eight are quiescent.  (C) Synchronous oscillations of all neurons. (D) All neurons at high voltage. }
    \label{fig:splitting}
\end{figure*}

For the small values of $g_C$ (sharp sigmoid), the phase separation into firing and quiescent neurons is the only stable state. 
 For intermediate values of  $g_C$ we still observe this stable state in Fig.~\ref{fig:splitting}A. We also show activity separation into oscillating and quiescent subnetworks in Fig.~\ref{fig:splitting}B. 
 Continuing to increase $g_C$ we obtain synchronous oscillation of the whole network in Fig.~\ref{fig:splitting}C and, finally, fixed point with constant uniform activity in Fig.~\ref{fig:splitting}D.

\subsection{Step function limit: All-to-all networks}
\label{subsec:step-all}

To better understand activity phase separation on the network, it is useful to consider a non-physiological limit of the model in which the 
sigmoidal functions describing both the firing rate and the dendritic adaptation are taken to be infinitely sharp, {\em i.e.},
step functions: $g_V = g_C = 0$.   In this case, neurons with above-threshold voltage $V^*$ fire at the 
maximal rate $r_m$, while neurons below that threshold voltage fire at the basal rate $r_b$. If the number of high and low firing rate neurons are  $n_h$ and  $n_l$ 
respectively ($n_h + n_l = N$ ) we find that

 \begin{eqnarray}
     V_h &=&  V_{eq} + \Delta V(C_h) \tau _V \left[(n_h - 1) r_m + n_l r_b \right]\\
     C_h &=& C_{eq} + \Delta C \tau _C \left[(n_h - 1) r_m + n_l r_b\right]. 
\end{eqnarray}
Similarly, the low firing-rate neurons have 
\begin{eqnarray}
     V_l &=&  V_{eq} + \Delta V(C_l) \tau _V \left[(n_l - 1) r_b + n_h r_m \right] \\
     C_l &=& C_{eq} + \Delta C \tau _C \left[(n_l - 1) r_b + n_h r_m\right].
\end{eqnarray}
One can check that rate of spikes received by a high firing-rate neuron $ R_{\rm high}=  \left[(n_h - 1) r_m + n_l r_b \right]$ is less than that received 
by a low firing-rate neuron $ R_{\rm low} = \left[(n_l - 1) r_b + n_h r_m \right]$.  However, the condition for being at a high firing rate is $V^* < V_h $ and a low firing rate is $V^* > V_l $.
For these inequalities to hold simultaneously with the result that $ R_{\rm high }<  R_{\rm low}$, one needs the high firing-rate neurons to be more sensitive to incoming spikes 
than the low firing-rate ones.  Thus we conclude that this state requires $C_l > C^* > C_h $.  
From this conclusion, we find $n_{l}$, the number of low-firing rate neurons, to be 
\begin{equation}
      n_l   = \left \lfloor \frac{(N r_m - r_b)\Delta C \tau _C + C_{eq} - C^*}{\Delta C \tau _C (r_m - r_b) } \right \rfloor,
      \label{eq:number-of-neurons} 
\end{equation}
where we have introduced the {\em floor} function: $\lfloor x \rfloor =$ the integer part of the real number $x$.

\begin{figure}
    \centering
    \includegraphics[width=85mm]{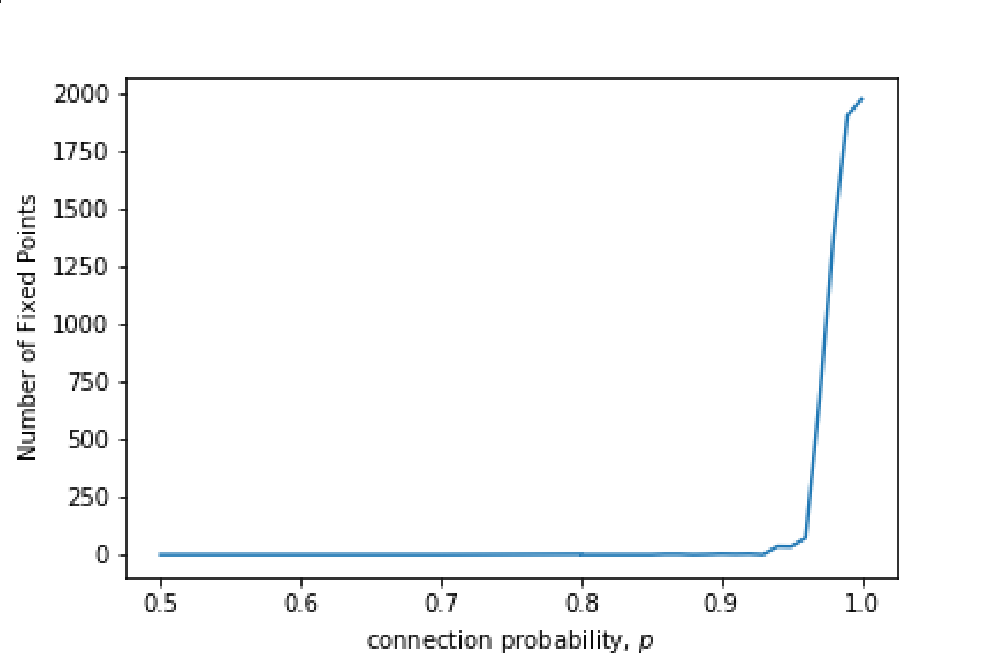}
    \caption{Number of stable fixed points as a function of the network connectivity probability $p$ for $N = 100$ neurons. For $p=1$ this number coincides with $\frac{n!}{n_l! n_h ! } $, and rapidly falls to one or zero when $p \lesssim 0.9$ . }
    \label{fig:numpoints}
\end{figure}

To observe the phase separated state, we require that the high-firing rate neurons remain sufficiently sensitive to incoming spikes.  The lower bound of
their sensitivity $\Delta V(C_{h})$ is given by 
\begin{equation}
  V^* -  V_{eq} < \Delta V(C_h) \tau _V \left[(n_h - 1) r_m + n_l r_b \right].
\end{equation}


While the number of neurons at low firing-rate neurons $n_{l}$ is fixed by Eq.~\ref{eq:number-of-neurons}, the identity of these neurons is determined solely by the initial conditions
on the all-to-all network. There are a large number $\frac{n!}{n_l! n_h ! } $ of otherwise identical fixed points that are related by permutation symmetry of the network.  
If, however, the network is more sparsely connected and thus does not have this permutation symmetry, there are fewer fixed points, as is discussed in the following section.

\section{Symmetry breaking on sparse networks}
\label{sec:ER}

If we randomly remove edges from the all-to-all network, we break the permutation symmetry of the neurons, and produce an instance of a network selected 
from ensemble of ER networks with  probability $p<1$ of a directed connection between neurons.   This leads to a rapid 
reduction in the number of stable fixed points with decreasing $p$,  as shown in Fig.~\ref{fig:numpoints}.

\begin{figure}
    \includegraphics[width=85mm]{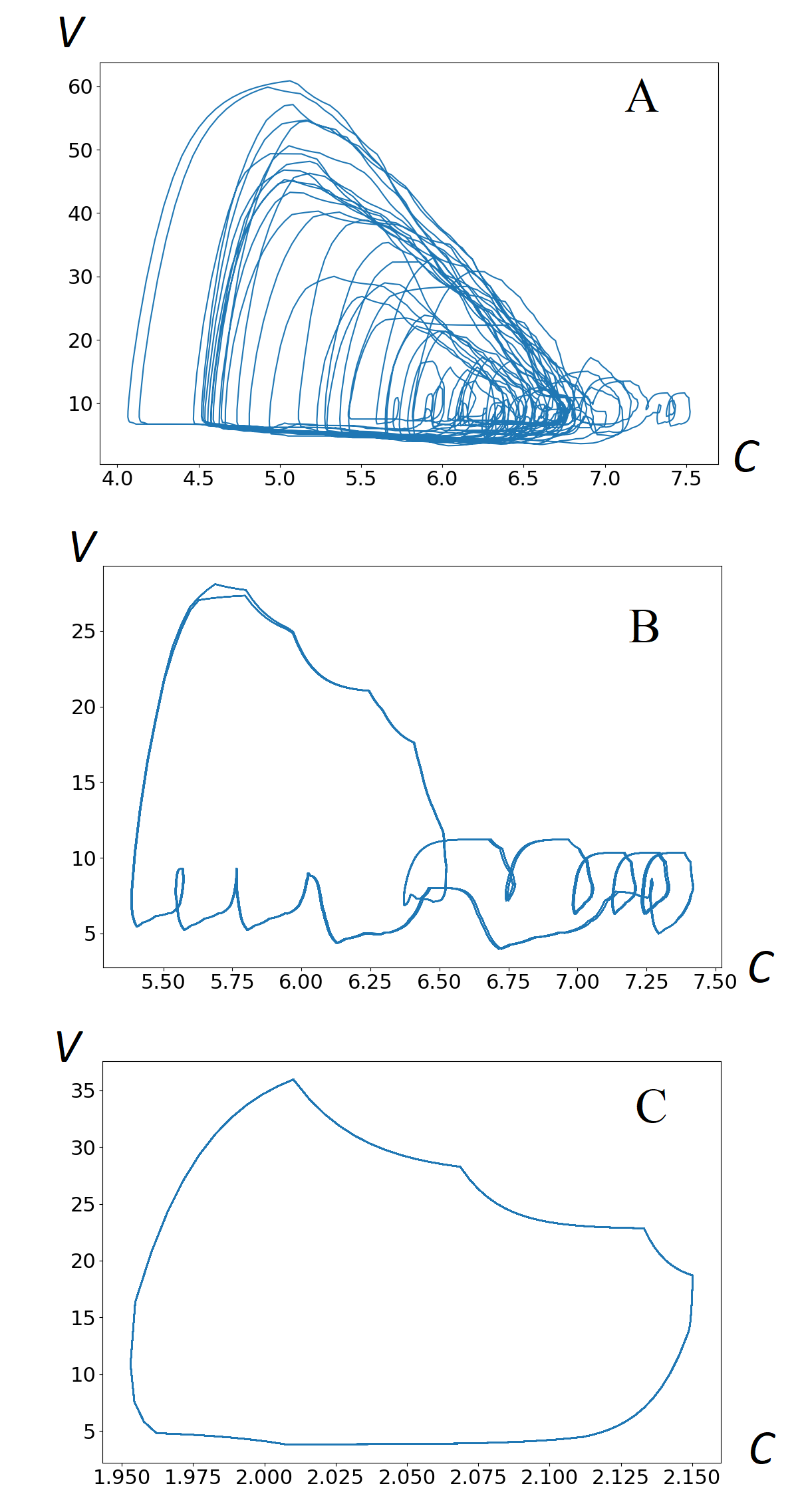}
    \caption{Phase trajectories of the network with step functions in the mean $V-C$ space. (A) Almost chaotic behavior. True chaos is not observed since the number of possible states is finite, but the voltage varies wildly. (B) Limit cycle with self-intersections, indicating asynchronous firing. (C) Standard limit cycle with synchronous firing, corresponding to true metronomic activity  (TMA), rarely observed for the case of step function limit.  }
    \label{fig:osc}
\end{figure}

Below $p \approx 0.9$ the number of stable fixed points drops to just one or vanishes entirely, resulting in only an oscillatory or approximately chaotic solution.  For the case of the step function neurons or 
for sufficiently sharp sigmoidal responses, we do not typically observe globally synchronized oscillations.  The asynchronous firing of different neurons, results 
in many self-crossing for the network averaged $V$ vs. $C$ graph, shown in Fig.~\ref{fig:osc}. 

While for the smooth sigmoids the most common case when oscillations occur is an unstable fixed point, it is not possible when sigmoids are very sharp. Indeed, any fixed point that not exactly on the threshold in this case is stable, as shown in Appendix~\ref{app:smoothness}. Therefore, the only opportunity for the oscillatory or approximately chaotic behavior is the absence of fixed point.   

\subsection{Oscillations on star networks}
\begin{figure}
    \centering
    \includegraphics[width=85mm]{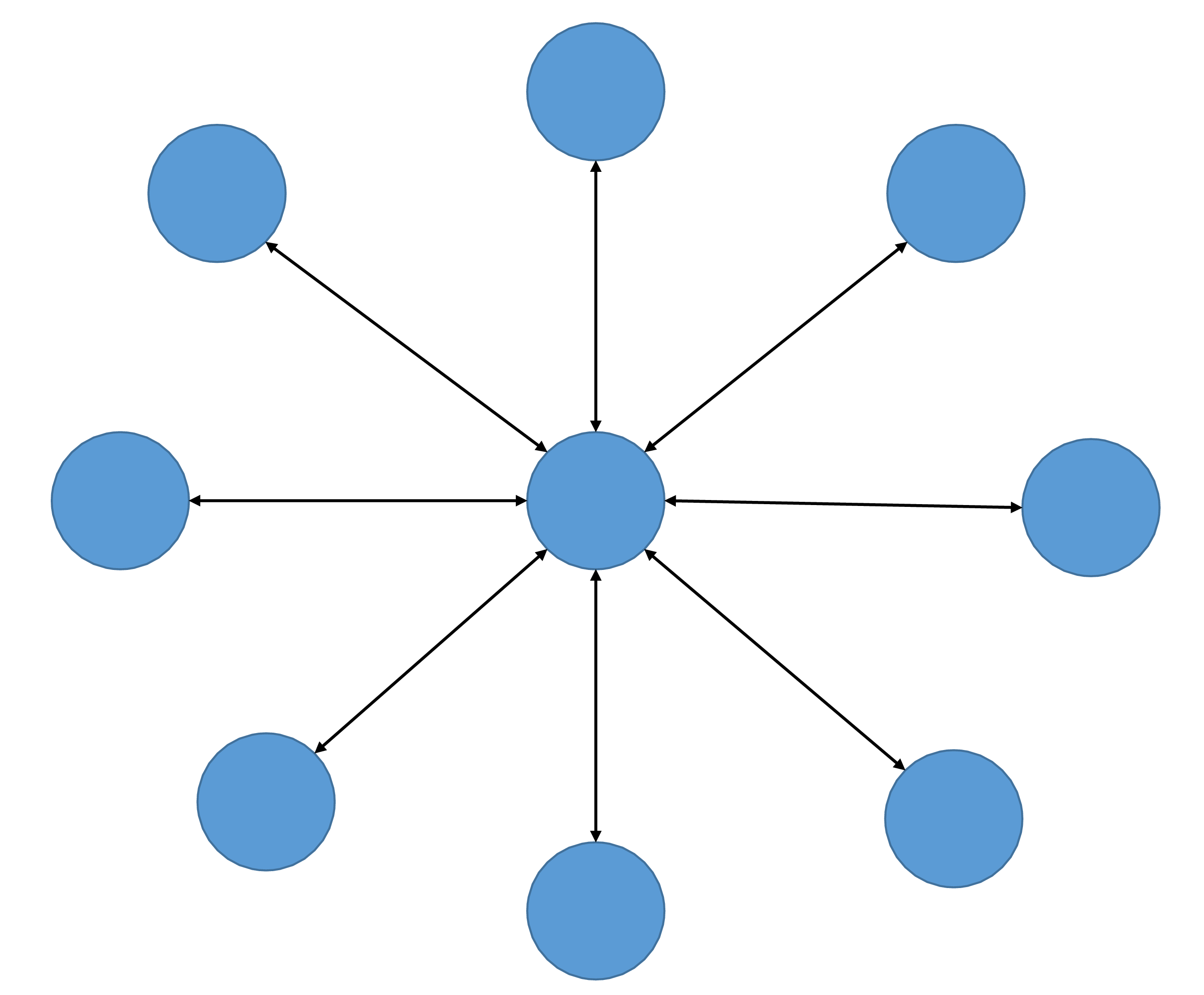}
    \caption{A star network with $N = 9 $ neurons. The peripheral neurons are bidirectionally coupled to the central neuron, but not to each other. }
    \label{fig:star}
\end{figure}

In order to understand how all fixed points can vanish for sparser networks, we can consider the special case of a star network, in which one central neuron is 
bidirectionally coupled to $N -1$ other neurons, and those other neurons are not coupled to each other. Such a network is shown in Fig.~\ref{fig:star}.  

We choose parameters such that the range of the single central neuron's firing rate is large enough to take the peripheral neurons across their firing-rate threshold:
\begin{equation}
\Delta V(0) \tau_V r_b < V^* < \Delta V(0) \tau_V r_m.
\end{equation}
Furthermore, we require that all the peripheral neurons firing together at their basal rate are able to excite the central neuron over threshold. If, however, the central neuron's dendritic calcium is above threshold, then all the peripheral neurons firing at their maximal rate are collectively insufficient to excite the
central neuron:
\begin{equation}
(N - 1 ) \Delta V (C > C^*) \tau_V r_m <  V^* < (N - 1) \Delta V(0) \tau_V r_b.
\end{equation}
 We also demand two conditions on the calcium threshold.  First, a single neuron cannot fire rapidly enough to push another neuron's dendritic calcium over threshold:
\begin{equation}
 C^* >  \Delta C \tau_C r_m.
 \end{equation}
$N-1$ neurons, however, firing at their maximal rate can induce calcium concentrations over threshold in the central neuron:
\begin{equation}
(N - 1) \Delta C \tau_C r_b < C^* < (N - 1) \Delta C \tau_C r_m,
\end{equation}
but they cannot do so when they are all firing at their basal rate. 

By obeying all of the above inequalities, the system cannot reach a fixed point.  Instead the network with these step function neurons oscillates. The 
central neuron excites the peripheral ones and then those neurons drive the central
neuron's calcium concentration above threshold rendering it insensitive.  As a result, the central neuron returns to its low firing state and then so do the peripheral ones.  At this point the cycle 
begins again.  

Recall, however, that the step-function neurons on the all-to-all coupled network do not oscillate, reach one of many fixed points characterized by activity phase separation. By breaking the permutation symmetry of the network, the star network admits a new synchronous oscillatory phase. This is reminiscent of an effect called converse symmetry breaking~\cite{Nishikawa:16}, where the necessary condition for synchronous activity of a coupled network of oscillators is an asymmetry of this system.  We observe a similar stability of globally 
synchronous oscillations in random networks that break the permutation symmetry such as the ER graphs discussed above.

\section{The effect of network heterogeneity on phase separation}
\label{sec:k-cores}

We have established that the neuron model leads generically to dynamical phase separation. On permutation symmetric all-to-all networks, this phase separation is a form of a spontaneously
broken symmetry, but it exists on more sparse networks too.  This poses the question: how does the network topology modify phase separation?  To address this, we consider another simplified 
limit of the model by setting $r_b = 0$ (i.e. a neuron is either firing or not), using a step function firing rate ($g_{V} \rightarrow 0$), and eliminating dendritic adaptation by taking 
$C^* \rightarrow \infty$. In this limit, we find phase separation into groups of firing and nonfiring neurons at the fixed point.  We demonstrate that the cluster of firing neurons is determined
exactly by the $k$-cores of the network for an integer $k$ determined by the remaining neuron parameters.

Consider $n_i$ actively firing input neurons synapsing on neuron $i$. From the fixed point condition $\dot{V}_i = 0$ we find
\begin{equation}
    \frac{V_i}{\tau _V} = n_i \Delta V r_{m}.
\end{equation}
For the $i^{\rm th}$ neuron to be part of the group of actively firing ones,  $V_i > V^*$,  which implies  that number $n_{i}$ of firing inputs exceeds a lower bound:
\begin{equation}
    n_i \geq \frac{V^*}{\tau _V \Delta V r_{max}}.
    \label{eq:n}
\end{equation}

We intend to relate the actively firing group with a topological feature of the network: a $k$-core.  This structure is defined to be the maximal subnetwork, such that within it 
each neuron has $k$ or more inputs from the other neurons in that subnetwork. $k$-cores have been discussed in a variety of applications in neuroscience, bioinformatics, ecology, and the study of social networks~\cite{Morone:19,Bader:03,Seidman:83,Lahav:16}.   The condition for a neuron to be in the actively firing group, Eq.~\ref{eq:n}, is equivalent to
membership within a $k$-core with the integer $k$ given by
\begin{equation}
    k = \left \lfloor \frac{V^*}{\tau _V \Delta V r_{\text{max}}} \right \rfloor.
\label{eq:k_predictor}
\end{equation}
If a $k$-core with some $k$ given by Eq. \ref{eq:k_predictor} is absent from the network, the dynamical system on that network relaxes to the quiescent fixed point, but
if such a $k$-core is present, the neurons making up the $k$-core become fixed in the HA phase.  We note that for typical values of $k$, most of the network will be part of that 
$k$-core~\cite{Dorogovtsev:06} because for $k > 2$ the probability of a neuron being part of the $k$-core has a discontinuous jump from zero to a significant value as a function of 
\begin{figure}[h]
    \centering
    \includegraphics[width=85mm]{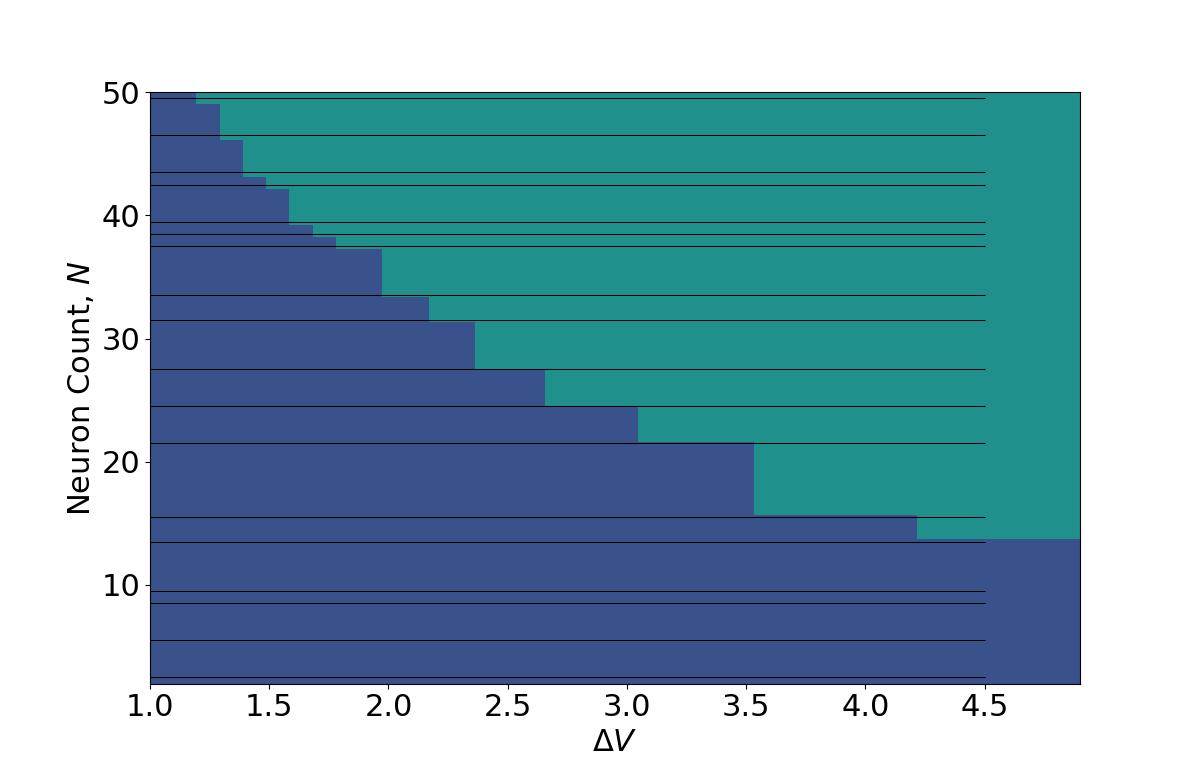}
    \caption{The phase diagram for the simplified model discussed in section~\ref{sec:k-cores}. There is no oscillatory phase, only quiescent (Q, blue) and high activity (HA, green). Black horizontal lines correspond to $k$-core transitions. We see almost exact correspondence between $k$-cores transtions and steps on the phase boundary. Small deviations are due to the fact that the average voltage of the whole network can be below $V^*$ even in the presence of the active $k$-core due to the averaging over all neurons including quiescent ones. }
    \label{fig:k_predictor}
\end{figure}
the density of synaptic connections (in the thermodynamic limit of large networks).   We compare our predicted phase boundary of the system and $k$-cores in Fig. \ref{fig:k_predictor}. 
The predicted $k$ value for a particular set of $n$ and $\Delta V$ parameters corresponds exactly to the point in the phase space where the HA phase gives way to the Q phase.
For this restricted version of the Feldman Del Negro model, at least, there is a precise correspondence between the neuronal network's dynamical phase behavior and 
the prediction made purely from the topology of the underlying network. $k$-cores
completely determine the dynamical phase transition of the neurons interacting on them, and how the network phase separates into groups of high and low activity neurons.

\section{Discussion}
\label{sec:Discussion}

We have explored the FDN model of oscillations in the  preB\"{o}tC and found a form of dynamical phase separation on the network in which groups of neurons separate into
high and low firing-rate fixed points.  This firing-quiescent phase separation plays the crucial role in the termination of the TMA phase. One feature that emerges from this work is 
that the permutation-symmetric system (the all-to-all coupled network) admits a type of spontaneous symmetry breaking into these high and low activity phases.  In more sparse, and 
physiologically relevant networks, the permutation symmetry is broken. In this case, the details of the network connectivity modify the inherent instability of the system 
toward phase separation into groups of high and low firing-rate neurons. In one particular limit of the model, we found that this interaction of neuronal dynamics and 
network topology is particularly simple.  By examining only the $k$-core structure of the network, one can precisely predict both the dynamical phase diagram and which neurons will 
end up in the high and low firing rate groups.

In the full model, the effect of the $k$-cores in determining the phase boundaries of the dynamical system remains, but no longer does it completely control these dynamics. The incomplete influence of $k$-cores was observed earlier~\cite{Schwab:10}. Here we believe we have better elucidated the underlying mechanism and explained why their control of the dynamics is not complete.

 To assess the importance of these observations for the physiological preB\"{o}tC, we first present the numerically computed phase 
diagram for 1000 neurons using parameters consistent with physiological measurements. This is shown in Fig.~\ref{fig:experiment}.  Please see appendix~\ref{app:parameters} for a 
discussion of how the neuronal and network parameters were selected.

As discussed in the appendix, there is a remaining uncertainty in determining the value of $\Delta C$. Moreover, the full preB\"{o}tC has 
somewhere between two to three times as many neurons as used in the simulation.  We note, however, a scaling argument, based on the mean-field analysis
of the model, that allows us to shift $\Delta C$ as a way of effectively changing the network's size.  In the mean-field theory,  three parameters $\Delta C$, $\Delta V$, 
and $p N$ appear in only two combinations $p N \Delta V $ and $p N \Delta C $.   As a result, if we change $\Delta C \rightarrow \frac{\Delta C}{\lambda}$, 
\begin{figure}[h]
    \includegraphics[width=85mm]{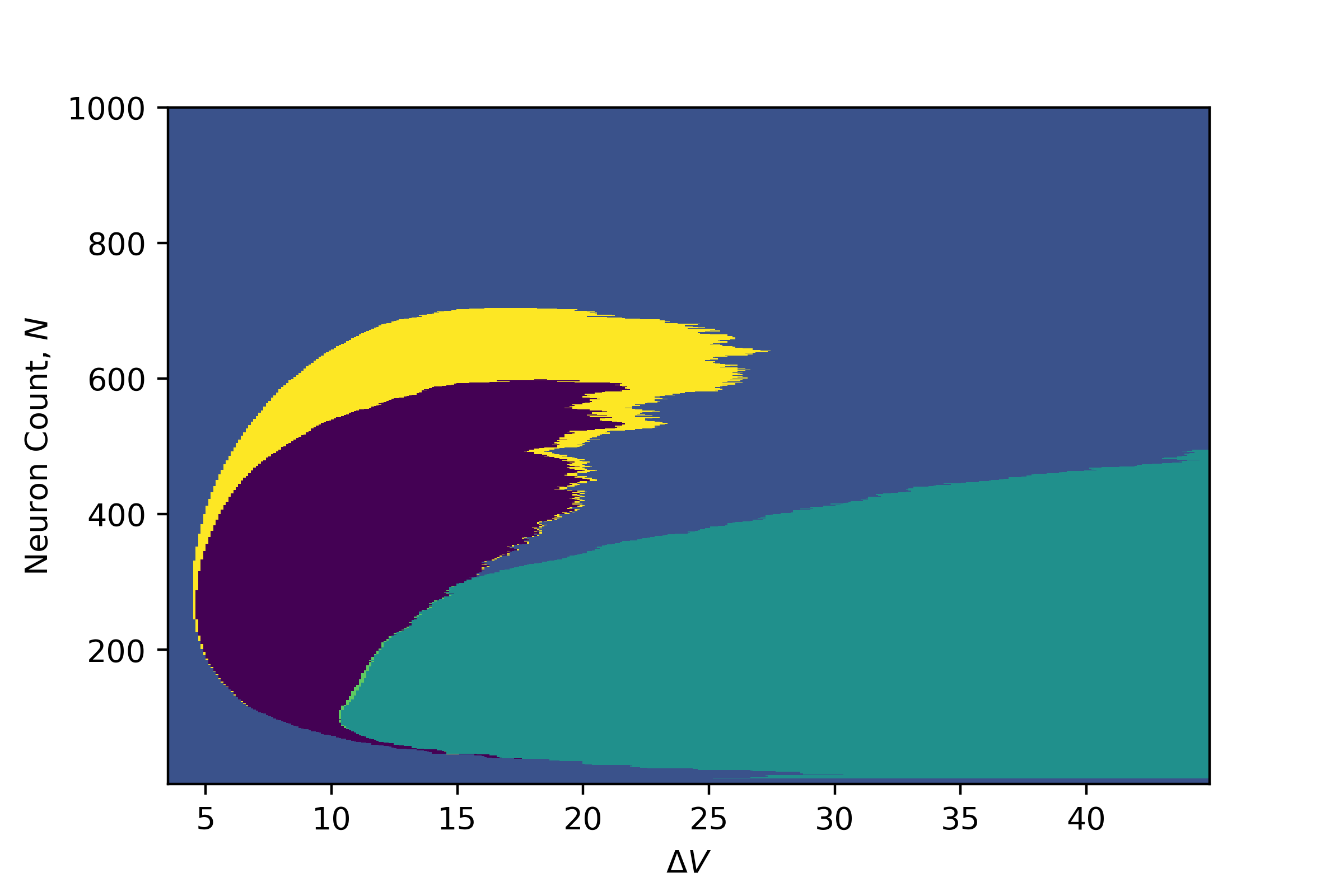}
    \caption{Phase diagram of large networks with $N$ up to 1000. All five phases are present: true metronomic activity (TMA) is purple, below threshold oscillations (BTO, yellow), above threshold oscillations (ATO, light green), high activity (HA, dark green), and quiescent (Q, blue).  The pattern is approximately the same as in figure \ref{fig:general}C, supporting the scaling argument.}
    \label{fig:full-size}
\end{figure}
$\Delta V \rightarrow \frac{\Delta V}{\lambda }$, and $p N \rightarrow p N \lambda$ the mean-field solutions are invariant.   We can test this scaling hypothesis in the full model by 
comparing the phase diagram of the N=1000 network with $\Delta C = 2.5 \times 10^{-3}$, shown in Fig.~\ref{fig:full-size}, to a much smaller network of $N=100$ and $\Delta C = 0.1$, 
shown in Fig.~\ref{fig:general}C.  Their correspondence supports out exploration of larger networks using calcium scaling.

The scaling hypothesis suggests that, if we were able to expand the network size used in Fig.~\ref{fig:experiment} to the preB\"{o}tC's true physiological size, we would find that the region of stable oscillations is bounded 
from above as well as for high and low neuronal excitability.  We see for Fig.~\ref{fig:full-size} that the large $N$ network has a rough phase boundary between the TMA and Q on the right side of the bounded
TMA domain, which is incompatible with the mean-field analysis and reflects the role of dynamical phase separation on the network.  This result makes an interesting prediction in that
there are regions of the phase diagram where {\em increasing} neuronal excitability can actually produce globally quiescent networks, through calcium inhibition.  This feature cannot
be reproduced by the mean-field model.  

Another consequence of this large $N$ phase diagram is that we can predict the robustness of the network to damage.  
From Fig.~\ref{fig:full-size}, we see that under optimal conditions, one can destroy about eighty percent of the network before causing the collapse of the oscillating phase.  This 
agrees with experimental observations~\cite{Gray:01}. 
\begin{figure}[h]
    \includegraphics[width=85mm]{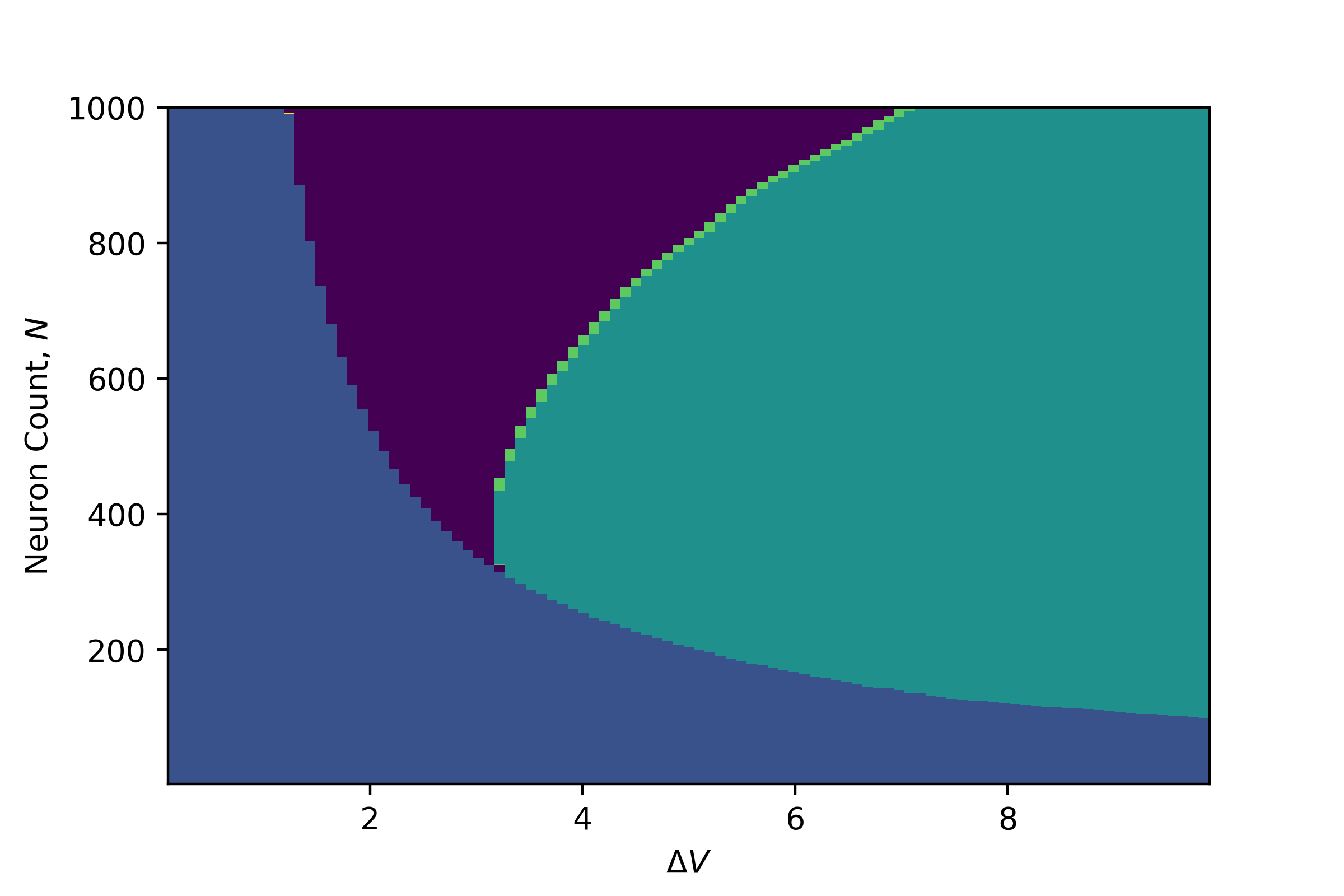}
    \caption{Phase diagram of the network with physiologically relevant parameters. It shows three 
    stable dynamical phases: true metronomic activity (TMA, purple) consistent with the preB\"{o}C's physiological 
dynamics, as well as a high activity (HA, dark green), and a quiescent (Q blue) regime. There is a narrow band of above threshold oscillations (ATO, light green).}
    \label{fig:experiment}
\end{figure}

We propose three types of experimental tests of the above analysis.  The first of these, alluded to above, is that the network should be able to be silenced by increasing neuronal 
excitability.  Secondly we predict that the roughness of the phase boundaries, particularly when $N$ is large, suggests presence of multiple reentrant transitions in which the network 
goes from being oscillatory to quiescent, and back to oscillatory as neurons are removed from it.  Third, one should be able to directly observe dynamical phase separation in the system. 
In either the high activity or quiescent phase, one should be able to find neurons trapped at the other fixed point so that the globally quiescent state of the network should harbor 
some fixed fraction of high firing rate neurons. Conversely, the network in its globally highly active state should contain a subpopulation of neurons trapped in their low firing-rate 
state.

\acknowledgements
AJL and VMS acknowledge partial support from NSF-DMR-1709785.  VMS acknowledges support from 
Peccei-Holmes Graduate Research Fellowship and the Bhaumik Institute for Theoretical Physics. AJL, VMS and MB are thankful to Jack Feldman, Robijn Bruinsma, 
and Sufyan Ashhad for fruitful discussions.

\appendix
\section{Determining physiological parameters for the model}
\label{app:parameters}
To prepare Fig.~\ref{fig:experiment} we must address the current understanding of the physiological parameters of the neurons as well as the network connectivity parameters. Most of them can
be fixed using experimental data shown in Table~\ref{tab:parameters}.  The exception is the set of parameters related to the dendritic calcium concentration, that are not currently as well known. We chose 
them to reproduce the observed dynamics of the system.

\begin{figure}[H]
    \includegraphics[width=85mm]{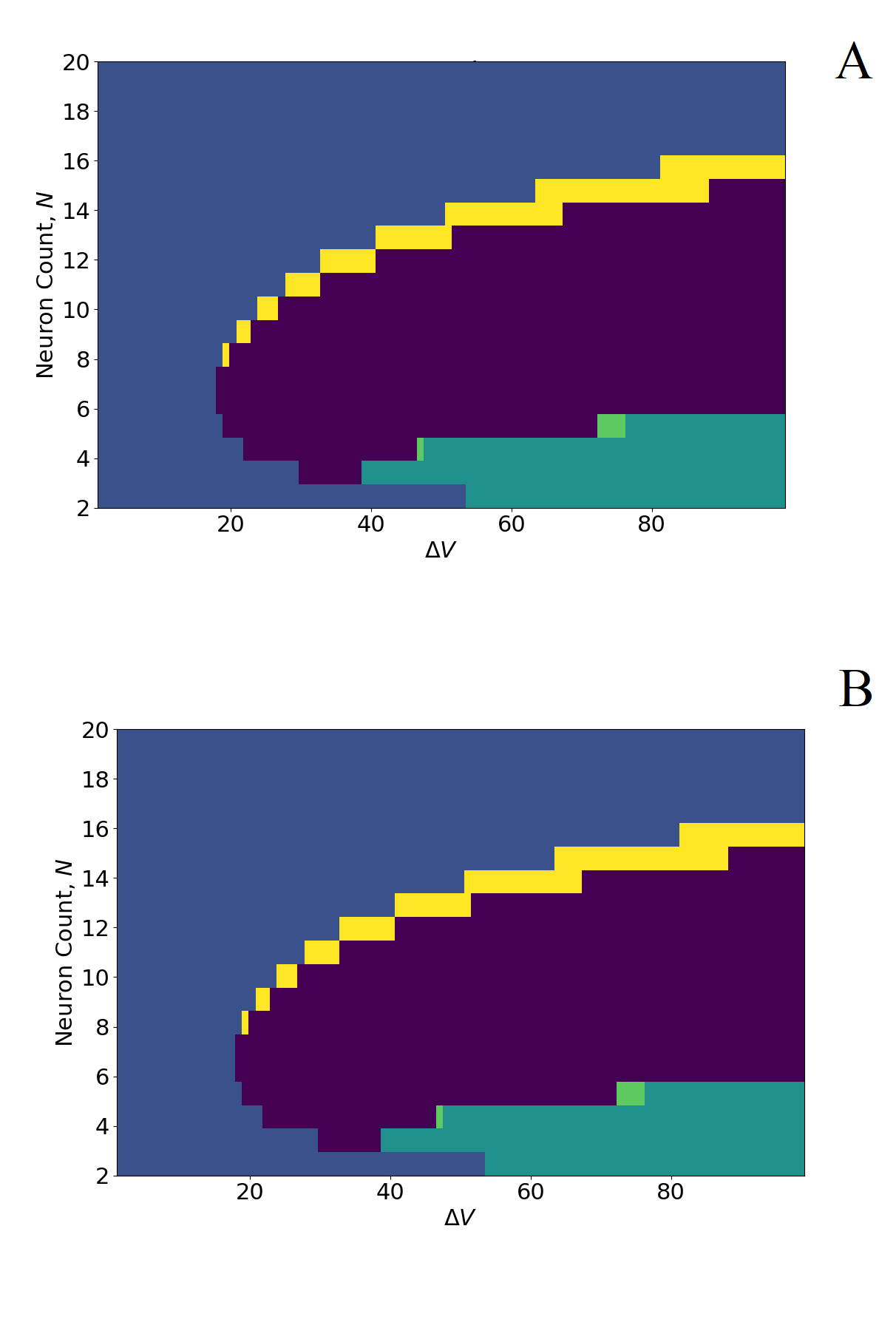}
    \caption{(A) Phase diagram for the all-to-all coupled network, using arbitrary initial conditions and smooth sigmoids. It is identical to (B), the mean-field phase diagram. }
    \label{fig:meanfield}
\end{figure}

\begin{table*}
\centering
\caption{Parameters used in the simulation}
\begin{tabular}{l|c|c|c|c|c|c|c|c|c|c|c|r}
\textbf{Fig.} & $g_C$ & $g_V$, mV & $C^*$ & $V^*-V_{eq}$,mV & $\tau_C$,ms & $\tau_V$,ms & $r_m$,Hz & $r_b$,Hz  & $\Delta C $ & $\Delta V_{max}$,mV  & $p$ & $N$  \\
\hline \hline
\ref{fig:general}A & 3 & 5 & 5 & 15 & 500 & 10 & 75 & 5 & 0.1 & 0-100 & 1 & 2-20
\\
\ref{fig:general}B & 0.01 & 5 & 5 & 15 & 500 & 10 & 75 & 5 & 0.1 & 0-100 & 1 & 2-20
\\
\ref{fig:general}C & 3 & 5 & 5 & 15 & 500 & 10 & 75 & 5 & 0.1 & 0-100 & 0.2 & 2-100
\\
\ref{fig:nonmon} & 5 & 5 & 15 & 15 & 500 & 10 & 75 & 5 & 0.035 & 0-25 & 0.75 & 2-100
\\
\ref{fig:nonmonotonicity} & 0 & 0 & 15 & 15 & 500 & 10 & 75 & 5 & 0.1 & 5-30 & 0.5 & 2-20
\\
\ref{fig:four-networks} & 0 & 0 & 15 & 15 & 500 & 10 & 75 & 5 & 0.1 & 10 & 0.5 & 6-9
\\
\ref{fig:nullclines}A & 0.3 & 0.5 & 5 & 15 & 500 & 10 & 75 & 5 & 0.1 & 50 & 1 & 10
\\
\ref{fig:nullclines}B & 3 & 5 & 5 & 15 & 500 & 10 & 75 & 5 & 0.1 & 50 & 1 & 10
\\
\ref{fig:splitting}A & 0.3 & 0.5 & 5 & 15 & 500 & 10 & 75 & 5 & 0.1 & 50 & 1 & 10
\\
\ref{fig:splitting}B & 0.3 & 0.5 & 5 & 15 & 500 & 10 & 75 & 5 & 0.1 & 50 & 1 & 10
\\
\ref{fig:splitting}C & 1.1 & 0.1 & 5 & 15 & 500 & 10 & 75 & 5 & 0.1 & 50 & 1 & 10
\\
\ref{fig:splitting}D & 10.8 & 1.8 & 5 & 15 & 500 & 10 & 75 & 5 & 0.1 & 50 & 1 & 10
\\
\ref{fig:numpoints} & 0 & 0 & 20 & 15 & 500 & 10 & 70 & 5 & 0.015 & 7.3 & 0.5-1 & 100
\\
\ref{fig:k_predictor} & 0 & 0 & $\infty$ & 15 & 500 & 10 & 70 & 5 & 0.1 & 1-5 & 0.5 & 1-50
\\
\ref{fig:full-size} & 3 & 5 & 5 & 15 & 500 & 10 & 75 & 5 & 0.025 & 1-50 & 0.083 & 1-1000
\\
\ref{fig:experiment} & 3 & 5 & 5 & 15 & 500 & 20 & 40 & 0.1 & 0.015 & 1-10 & 0.065 & 1-1000
\\
\ref{fig:meanfield}A & 3 & 5 & 5 & 15 & 500 & 10 & 75 & 5 & 0.1 & 0-100 & 1 & 2-20
\\
\ref{fig:transparent}A & 0-3 & 0-20 & 5 & 15 & 500 & 10 & 75 & 5 & 0.1 & 0-100 & 1 & 2-20
\end{tabular}
\label{tab:fig-par}
\end{table*}

 Specifically, we set $\tau_C$ to reproduce the observed period of stable oscillation. Taking into account that, for us, the 
units for the calcium concentration are arbitrary as is the choice of the zero for that concentration, we are left with only two independent parameters: $g_C$ and $\Delta C$. 
$g_C$ is chosen to be large enough that we have reproducible phase behavior, avoiding highly heterogeneous and initial-condition dependent results as 
shown in Fig.~\ref{fig:general}B.  We also require it to be small enough to produce a true threshold for calcium inactivation of the dendrite, {\em i.e.}, $ \frac{C_{eq} - C^*}{g_C} > 1 $. 
The last parameter to be fixed is  $\Delta C$. The proper choice of $\Delta C$ is facilitated by scaling behavior observed in the 
mean-field approximation of the model. 

Indeed, consider Fig.~\ref{fig:general}C. To quantitatively fit the {\em in vitro} data, the network must support 
stable oscillations when $\Delta V \approx 2.8 $ mV, which is the average magnitude of an EPSP~\cite{Rekling2000}.   To do that, we choose $\Delta C = 0.007 $ and find, with no remaining 
fitting parameters, that stable oscillations occur for networks $N \approx 10^3 $ neurons --  see Figure \ref{fig:experiment}. It is computationally difficult to study networks with
$N > 2000$, but we can use the rescaling property, observed in the mean-field approximation, to qualitatively predict the behavior for larger $N$. If we 
choose $\Delta C = 0.025, p = 0.083$, we obtain the phase diagram, shown in Fig.~\ref{fig:full-size}. The result is quite similar to that shown in Fig.~\ref{fig:general}C, which 
describes on a much smaller system. 

\section{Mean-field solution}
\label{sec:mean-field}

The mean field solution is obtained  by assuming that somatic potentials and dendritic calcium concentrations of all the neurons are the same: $V_i = V, C_i = C$. Another assumption is that the network is on average homogeneously connected,  {\em i.e.} each neuron on average has $p N $ inputs and outputs, with $N$ the total number of neurons and $p$ the connection probability. Then Eqs.~\ref{eq:V}, \ref{eq:C} are rewritten as pair of equations for $V$ and $C$

\begin{eqnarray}
    \label{eq:Vmean}
    \frac{dV}{dt} &=& \frac{1}{\tau _V}(V_{eq}-V) + \Delta V(C) p (N-1) r(V)\\
     \frac{dC}{dt} &=& \frac{1}{\tau _C}(C_{eq}-C) + \Delta C p (N-1) r(V),
    \label{eq:Cmean}
\end{eqnarray}

\begin{figure}[h]
    \includegraphics[width=85mm]{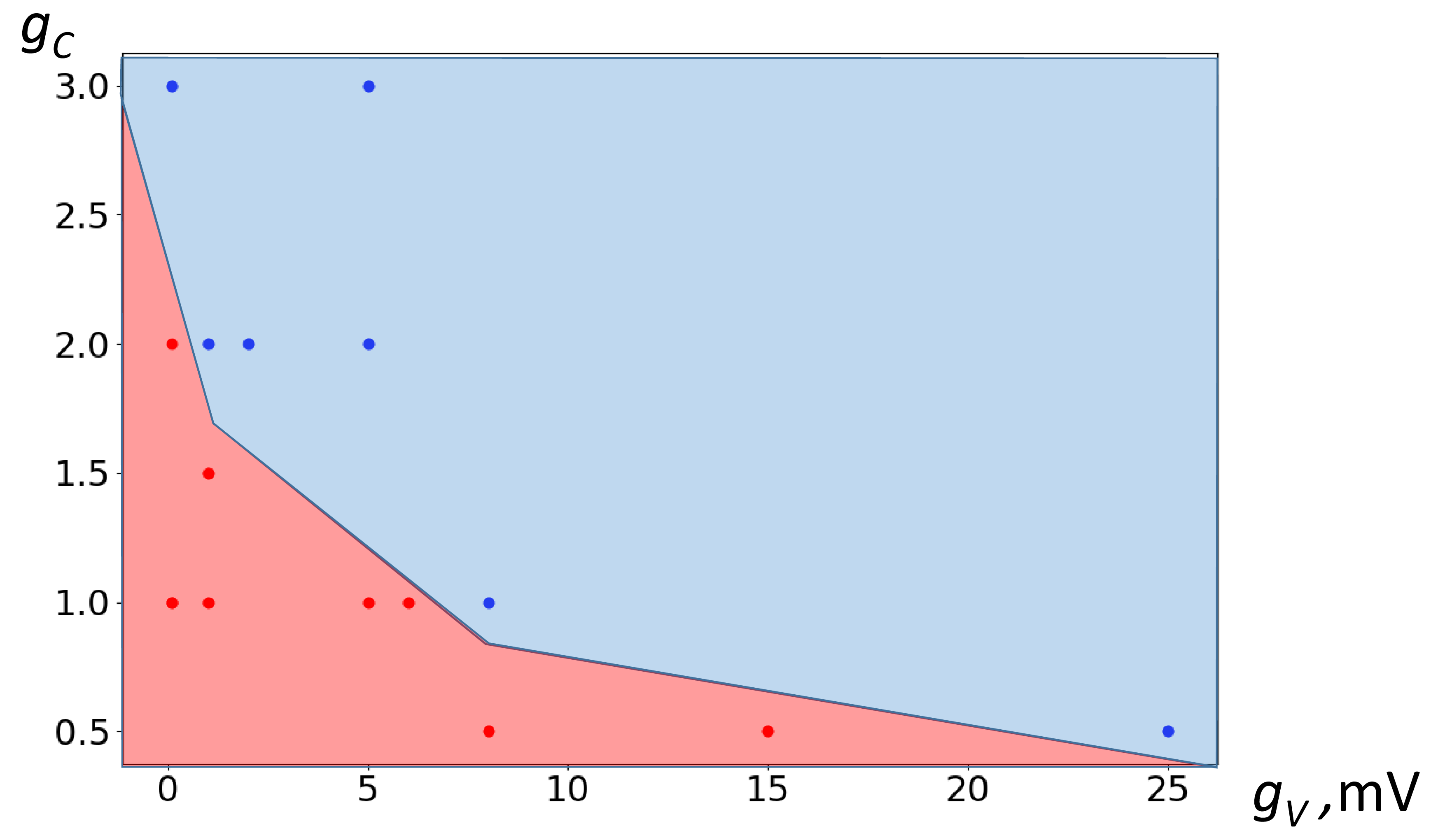}
    \caption{The results of the simulations for different $g_V$ and $g_C$. Blue points corresponds to the case when these results (phase diagrams) fit mean-field approximation, and red ones to the cases when they do not (we disstinguish it by visually comparing corresponding phase diagrams).  Transparent regions are guides for the eye.  }
    \label{fig:transparent}
\end{figure}

The phase diagram for such system is shown in Fig.~\ref{fig:meanfield}B. One may see that it is identical to the phase diagram for the all-to-all coupled network in Fig.~\ref{fig:meanfield}A in case of smooth transition in dendritic sensitivity. The mean-field approximation is valid for large $g_V$ and $g_C$ but breaks when they become smaller. More detailed results are shown in Figure~\ref{fig:transparent}. 

\section{Stability of fixed points on sparse networks}
\label{app:smoothness}

\begin{figure}[h]
    \includegraphics[width=85mm]{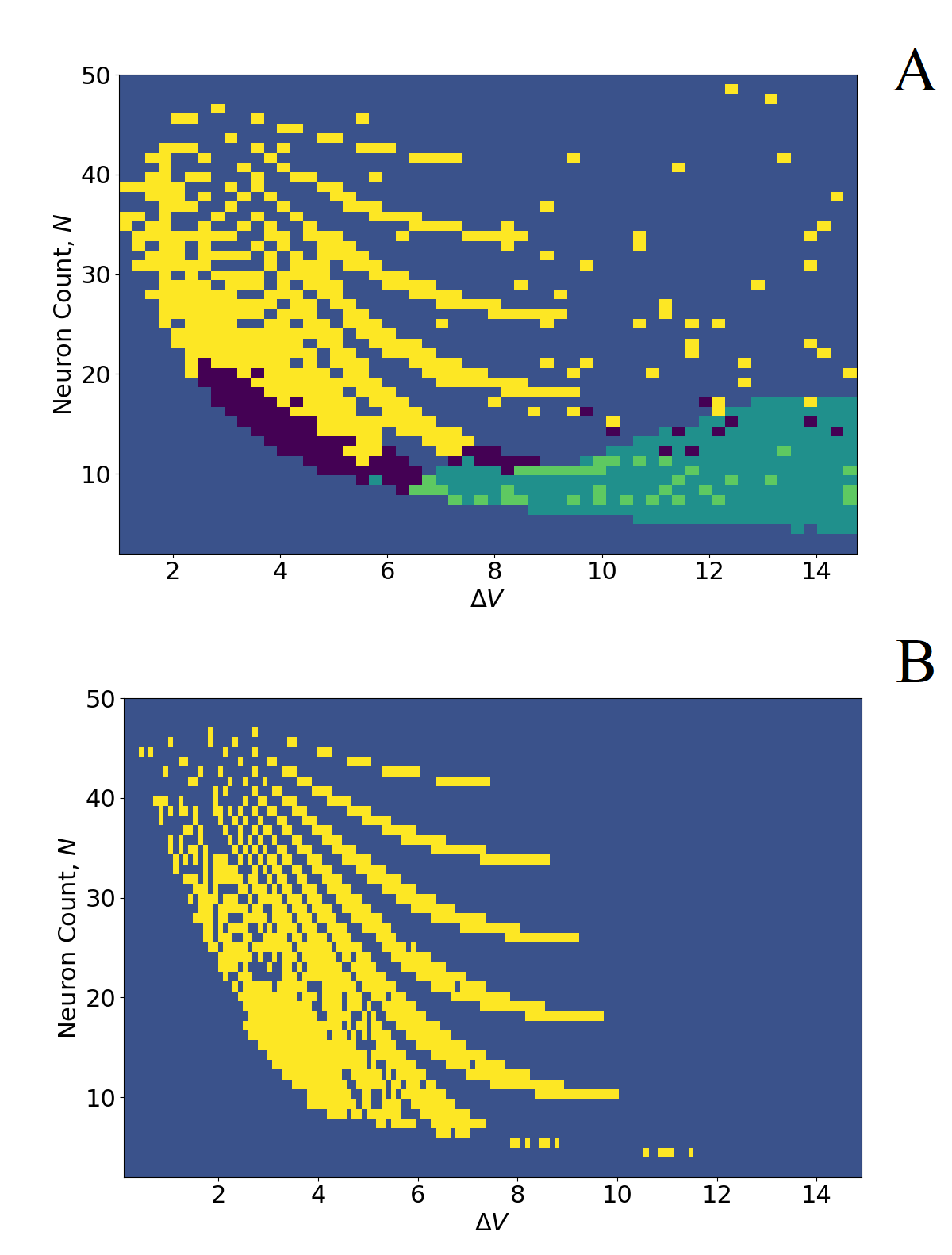}
    \caption{(A) Phase diagram for the all-to-all coupled network, $g_C = 0, g_V > 0$. All five previously mentioned phases are present. There is a quasi-periodical pattern on the BTO-Q boundary (yellow and blue). This phase diagram fits the theoretical prediction (B) where blue corresponds to the case that Eq.~\ref{eq:single} has a solution and yellow to the case that it does not.}
    \label{fig:meanfield}
\end{figure}

For the smooth sigmoid neurons, we found in section~\ref{sec:fully}  that there was only a single fixed point -- see Fig.~\ref{fig:nullclines}.  When that fixed point is unstable, the system executes
limit cycle oscillations.  To determine the parameter range of these oscillations,  here we investigate the stability of that fixed point. Expanding the equations of motion near the 
fixed point $\{ V_i^f, C_i^f\}$ in the $2N$ dimensional space of $V_{i},C_{i}$, we define $v_i = V_i - V_i^f, c_i = C_i - C_i^f $ and obtain 
\begin{eqnarray}
\nonumber
    \frac{dv_i}{dt} &=& -\frac{v_i}{\tau _V} +
    \Delta V'(C_i^f) c_i \sum _{j } M_{i j} r(V_j^f) + \\
    & & + \Delta V(C_i^f) \sum _{j } M_{i j} r'(V_j^f) v_j\\
\frac{dc_i}{dt} &=& -\frac{c_i}{\tau _C} + \Delta C \sum _{j } M_{i j}  r'(V_j^f) v_j
\end{eqnarray}
Using $r(V)$ and $\Delta V(C)$  from Eqs.~\ref{eq:VC},~\ref{eq:rV}, we find

 \begin{eqnarray}
 \nonumber
    \frac{dv_i}{dt} & & = -\frac{v_i}{\tau _V} - \frac{1}{g_C} \Delta V_{max} \sigma'(\frac{C^* - C_i^f}{g_C}) c_i \sum _{j } M_{i j} r(V_j^f) + 
    \\
    & & +  \frac{1}{g_V} (r_m - r_b)  \Delta V(C_i^f) \sum _{j } M_{i j} \sigma'(\frac{V_j^f - V^*}{g_V}) v_j,
\label{eq:stabV}
\end{eqnarray}

and
\begin{equation}
    \frac{dc_i}{dt} = -\frac{c_i}{\tau _C} + \Delta C (r_m - r_b)  \sum _{j } M_{i j}  \sigma'(\frac{V_j^f - V^*}{g_V})v_j.
    \label{eq:stabC}
\end{equation}
Dynamical phase separation requires that neither $C^* - C_i^f $ nor $V^* - V_i^f $ vanish. Therefore, if $g_V$ and $g_C$ are small, the terms with sigmoids are 
exponentially suppressed. Neglecting these we see that only the first term on the right hand side of Eqs.~\ref{eq:stabV} and~\ref{eq:stabC} remains. These imply stability of the fixed point. For large $g_C$ and $g_V$, however, we cannot ignore the terms proportional to $\sigma'$, 
which destabilize the phase separated fixed point.

\section{Quasi-periodical phase diagrams}
\label{app:fractal}

We consider the case $g_C = 0, g_V > 0 $, where quasi-periodical phase diagram can emerge.  Following subsection~\ref{subsec:step-all}  assuming the activity phase separation to $n_h$ neurons with $V_h, C_h$ and $n_l$ neurons with $V_l, C_l$ we write equations for the fixed point 

\begin{eqnarray}
\label{eq:Vh}
     V_h -  V_{eq} &=&  \Delta V(C_h) \tau _V \left[ (n_h - 1) r(V_h) + n_l r(V_l) \right] \\
     C_h - C_{eq} &=&  \Delta C \tau _C \left[ (n_h - 1) r(V_h) + n_l r(V_l) \right]. 
    \label{eq:Ch}
\end{eqnarray}
and
\begin{eqnarray}
\label{eq:Vl}
     V_l -  V_{eq} &=&   \Delta V(C_l) \tau _V \left[ (n_l - 1) r(V_l) + n_h r(V_h) \right] \\
     C_l - C_{eq} &=&  \Delta C \tau _C \left[ (n_l - 1) r(V_l) + n_h r(V_h) \right].
     \label{eq:Cl}
\end{eqnarray}

 In case  $g_C = 0$ we have $\Delta V(C_l) = 0, \Delta V(C_h) = \Delta V_{max}$ (for $ C_h < C^*, C_l > C^* $ . Then in the same way as in the subsection~\ref{subsec:step-all} we obtain  the number of neurons firing at low rate

\begin{equation}
      n_l   = \left \lfloor \frac{(N r(V_h) - r(V_l))\Delta C \tau _C + C_{eq} - C^*}{\Delta C \tau _C (r(V_h) - r(V_l)) } \right \rfloor,
      \label{eq:number-ofneurons} 
\end{equation}
and simplify equations for the somatic potentials to  get $V_l = V_{eq} $ and

 \begin{equation}
   \frac{V_h -  V_{eq}}{\Delta V_{max} \tau _V } =   N r(V_l) - r(V_h)  
   + \left \lceil \frac{\frac{C^*}{\Delta C \tau_C} - (N - 1)  r(V_l)  }{r(V_h) - r(V_l) } \right \rceil 
 \label{eq:single} 
 \end{equation}
where the $ceil$ function $\lceil x \rceil $ is the smallest integer that is larger or equal to $x$. 
 
 Eq.~\ref{eq:single} does not have a solution for $V_h$ for some parameters. Due to the ceil function on the right hand side, the changes happen when its argument is incremented by one, what causes quasi-periodical structure of the phase diagram.

\section{Simulation details}
\label{app:fig-parameters}

Source code is available at https://github.com/ mbibireata/Networks. For the simulation, we use the parameters from Table \ref{tab:fig-par}.


\bibliography{bib_kei.bib}

\end{document}